\documentclass[preprint,jcp,floats,amsmath,amssymb]{revtex4}
\usepackage{epsfig}

\usepackage{graphicx}
\usepackage{dcolumn}
\usepackage{bm}
\usepackage{chemarr}
\usepackage{multirow}
\usepackage{subfigure}
\usepackage{epstopdf}

\DeclareGraphicsExtensions{.png,.gif,.jpg,.pdf,.bmp}

\begin{document}

\title{How accurate are the non-linear chemical Fokker-Planck and chemical Langevin equations?}

\author{Ramon Grima $^{1,\footnotemark[2]}$, Philipp Thomas $^{1,2,\footnote[2]{Authors contributed equally to this work}}$, Arthur V. Straube$^{2}$}

\affiliation{$^{1}$ School of Biological Sciences, University of Edinburgh, UK \\ $^{2}$ Department of Physics, Humboldt University of Berlin, Germany}

\begin{abstract}
The chemical Fokker-Planck equation and the corresponding chemical Langevin equation are commonly used approximations of the chemical master equation. These equations are derived from an uncontrolled, second-order truncation of the Kramers-Moyal expansion of the chemical master equation and hence their accuracy remains to be clarified. We use the system-size expansion to show that chemical Fokker-Planck estimates of the mean concentrations and of the variance of the concentration fluctuations about the mean are accurate to order $\Omega^{-3/2}$ for reaction systems which do not obey detailed balance and at least accurate to order $\Omega^{-2}$ for systems obeying detailed balance, where $\Omega$ is the characteristic size of the system. Hence the chemical Fokker-Planck equation turns out to be more accurate than the linear-noise approximation of the chemical master equation (the linear Fokker-Planck equation) which leads to mean concentration estimates accurate to order $\Omega^{-1/2}$ and variance estimates accurate to order $\Omega^{-3/2}$. This higher accuracy is particularly conspicuous for chemical systems realized in small volumes such as biochemical reactions inside cells. A formula is also obtained for the approximate size of the relative errors in the concentration and variance predictions of the chemical Fokker-Planck equation, where the relative error is defined as the difference between the predictions of the chemical Fokker-Planck equation and the master equation divided by the prediction of the master equation. For dimerization and enzyme-catalyzed reactions, the errors are typically less than few percent even when the steady-state is characterized by merely few tens of molecules.
\end{abstract}

\maketitle

\section{Introduction}

Chemical master equations (CMEs) are the accepted mathematical description of chemical systems in well-mixed conditions \cite{Gillespierev}. These equations provide a mesoscopic description of chemical kinetics, interpolating between the microscopic regime of molecular dynamics and the macroscopic regime of rate equations (REs). It has been shown that CMEs are exact descriptions for any well-stirred and thermally equilibrated gas-phase chemical system \cite{Gillespie1992}. More recently it has been rigorously confirmed that their validity extends to chemical reactions in well-stirred dilute solutions \cite{Gillespie2009}. However, well before these rigorous demonstrations of the microscopic physical basis of the CME, scientists have employed these equations to probe the nature of mesoscopic chemical kinetics and in particular to understand how this may differ from kinetics on macroscopic length scales (see McQuarrie for a review \cite{McQuarrie} of the literature up till 1967). 

We briefly review the CME formalism. Consider a general chemical system consisting of a number $N$ of distinct chemical species interacting via $R$ elementary chemical reactions of the type:
\begin{equation}
s_{1j} X_1 + ... + s_{Nj} X_{N} \xrightarrow{k_j} \ r_{1j} X_{1} + ... + r_{Nj} X_{N}.
\end{equation}
Here $j$ is an index running from 1 to $R$, $X_i$ denotes chemical species $i$, $s_{ij}$ and $r_{ij}$ are the stoichiometric coefficients and $k_j$ is the macroscopic rate of reaction. If this system is well-mixed then its mesoscopic state is fully determined by the vector of the absolute number of molecules of each species, $\vec{n}=(n_1,...,n_N)^T$, where $n_i$ is the number of molecules of the $i^{th}$species. The CME is then a time-evolution equation for the probability of the system being in a particular mesoscopic state \cite{Gillespierev,vanKampen}:  
\begin{equation}
\frac{\partial P(\vec{n},t)}{\partial t} = \Omega \sum_{j=1}^{R} \biggl( \displaystyle\prod_{i=1}^N E_i^{-S_{ij}} - 1 \biggr) \hat{f}_j(\vec{n},\Omega) P(\vec{n},t),
\end{equation}
where $\Omega$ is the volume of the compartment in which the reactions occur and $E_i^{x}$ is a step operator -- when it acts on some function of the absolute number of molecules, it gives back the same function but with $n_i$ replaced by $n_i + x$. The chemical reaction details are encapsulated in the stoichiometric matrix $S_{ij}=r_{ij}-s_{ij}$ and in the microscopic rate functions $\hat{f}_j(\vec{n},\Omega)$. The probability that the $j^{th}$ reaction occurs in the time interval $[t,t+dt)$ is given by $\Omega \hat{f}_j(\vec{n},\Omega) dt$.  For elementary reactions, the microscopic rate function takes one of four different forms, depending on the order of the $j^{th}$ reaction: (i) a zeroth-order reaction by which a species is input into a compartment gives $\hat{f}_j(\vec{n},\Omega)=k_j$; (ii) a first-order unimolecular reaction involving the decay of some species $h$ gives $\hat{f}_j(\vec{n},\Omega)=k_j n_h \Omega^{-1}$; (iii) a second-order bimolecular reaction between two molecules of the same species $h$ gives $\hat{f}_j(\vec{n},\Omega)= k_j n_h (n_h-1) \Omega^{-2}$; (iii) a second-order bimolecular reaction between two molecules of different species, $h$ and $v$, gives $\hat{f}_j(\vec{n},\Omega)=k_j n_h n_v \Omega^{-2}$. 

The RE description of the same system is much simpler. Denoting the macroscopic concentration of species $i$ by $\phi_i$, the set of REs describing the macroscopic kinetics of the reactive system represented by Eq. (1)  are given by:
\begin{equation}
\frac{\partial \phi_i}{\partial t} = \sum_{j=1}^{R} S_{ij} f_j(\vec{\phi}),
\end{equation}
where $\vec{\phi} = (\phi_1,...,\phi_N)^T$ is the vector of macroscopic concentrations and $f_j$ is the macroscopic rate function of the $j^{th}$ reaction which has the general mass-action form, $f_j(\vec{\phi}) = k_j \prod_{m=1}^N \phi_m^{s_{mj}}$. REs provide a continuous deterministic ``many molecule'' description of kinetics. This strongly contrasts with the CME description which constitutes a discrete, stochastic, ``any number of molecule'' description that is faithful to the underlying microscopic basis of chemical reactions. 

Unfortunately, one of the main advantages of CMEs over their RE cousins, their discrete description, is also the source of their computational intractability. Differential-difference equations, such as the CME \cite{McQuarrie}, do not lend themselves easily to analysis. In contrast, there is a vast body of literature in engineering, mathematics and physics dealing with the analysis and solution of differential and partial differential equations. Thus at an early stage, considerable effort was invested in obtaining a partial differential approximation of the CME. In the 1940's, Kramers \cite{Kramers} and Moyal \cite{Moyal} developed a Taylor series expansion of the CME; by assuming that all terms with derivatives greater than two are negligible, one obtains the chemical Fokker-Planck equation (CFPE, \cite{Gardiner}), a second-order partial differential equation of the form:
\begin{equation}
\frac{\partial P(\vec{n},t)}{\partial t} = \Omega \sum_{j=1}^{R} \biggl( -\sum_{i=1}^N S_{ij} \frac{\partial}{\partial n_i} + \frac{1}{2} \sum_{i,w=1}^N S_{ij} S_{wj} \frac{\partial^2}{\partial n_i \partial n_w} \biggr) \hat{f}_j(\vec{n},\Omega) P(\vec{n},t).
\end{equation}
As Gardiner mentions in his book \cite{Gardiner}, ``this procedure enjoyed wide popularity -- mainly because of the convenience and simplicity of the result'' and also because ``it is often simpler to use the Fokker-Planck equation than the Master equation.'' A major and important difference between the CME and the CFPE is that $n_i$ is a positive integer for the CME while it is a real number for the CFPE. 

Several authors have questioned the validity of the CFPE approximation. The approximation is obtained by a perfunctory truncation of the Taylor expansion and hence it appears to be an uncontrolled and unjustified approximation of the CME. van Kampen, in particular, was a leading and influential critic of the CFPE approximation. In the 1960's and 70's, he developed a systematic perturbative expansion of the CME in powers of the inverse square root of the system volume $\Omega$ (the system-size expansion) and used it to show that to lowest order in the expansion, i.e. in the limit of large volumes -- the macroscopic limit, one obtains a Fokker-Planck equation which is of a different form than the CFPE \cite{vanKampen1961,vanKampen1976}. Of particular concern is that van Kampen's Fokker-Planck equation is linear whereas the CFPE is non-linear. Note that by non-linear Fokker-Planck equation here we mean one such that its drift and diffusion coefficients are generally non-linear functions of the molecule numbers $n_i$; this convention is adopted since it is in mainstream use, for example see the book by van Kampen \cite{vanKampen}. Taking into account higher-order terms in the system-size expansion does not lead to the CFPE as well. However, interestingly, in the limit of large volumes, the CFPE does reduce to van Kampen's linear Fokker-Planck equation \cite{Gardiner}. This led van Kampen to conclude that any features arising from the non-linear character of the CFPE are spurious and not to be taken seriously \cite{vanKampen1982}. We note that the limit of large volumes in van Kampen's system-size expansion is taken at fixed macroscopic concentrations and hence it corresponds to the limit of large molecule numbers \cite{vanKampen}. Hence van Kampen's conclusions can be equivalently stated as: the CFPE becomes a legitimate approximation of the CME in the limit of large molecular populations.  

A few studies at the time \cite{Kurtz,Horsthemke} did suggest that the CFPE's validity extended beyond the linear regime. Of particular importance is a result of Horsthemke and Brenig \cite{Horsthemke} which motivated the present study. The authors considered a simple dimerization reaction $\O \rightarrow X, X + X \rightarrow Y$ whereby molecules of a monomer species $X$ are introduced in a compartment of volume $\Omega$ and subsequently they bind to each other to form dimers $Y$. Assuming stationary conditions, the CME and CFPE are solved exactly. It is shown that the average concentration of monomers and the variance of the fluctuations from the two formalisms agree exactly to order $\Omega^{-1}$ and are respectively equal to $\phi + (8 \Omega)^{-1}$ and $(3/4) \phi \ \Omega^{-1}$, where $\phi$ is the macroscopic concentration obtained by solving the corresponding RE in steady-state conditions. The same example can be found worked in van Kampen's book \cite{vanKampen} wherein he shows that the linear noise approximation gives mean and variance equal to $\phi$ and $(3/4) \phi \ \Omega^{-1}$. As we mentioned before, a linearization of the CFPE will lead to the linear-noise approximation and hence from this example we can conclude that the non-linearity of the CFPE is non-spurious since it leads to a more accurate concentration estimate than that which is obtained from the linear-noise approximation. However one could argue that this higher accuracy is only particular to the dimerization example and not a general feature of the CFPE. Because of this or other reasons, the results of Hortshemke and Brenig do not appear to have received the attention they deserved at the time and van Kampen's conclusions about the CFPE were accepted, by and large, by the statistical physics community.

Approximately 40 years later after the inception of the system-size expansion, Gillespie revived the question of the validity of the CFPE by deriving  it without invoking truncation of the Kramers-Moyal expansion of the CME \cite{Gillespie2000}. To be precise, he derived the chemical Langevin equation (CLE):
\begin{align} 
\frac{\partial}{\partial t} n_i(t)=\Omega \sum_{j=1}^{R} S_{ij} \hat{f}_j(\vec{n}(t),\Omega)+\Omega^{1/2} \sum_{j=1}^{R} S_{ij} \sqrt{\hat{f}_j(\vec{n}(t),\Omega))} \Gamma_j(t),
\end{align} 
where $\Gamma_j(t)$ are temporally uncorrelated, independent Gaussian white noises. This stochastic differential equation is exactly equivalent to the CFPE in the sense that its solution  generates exact sample paths of the CFPE, Eq. (4). Essentially he showed that the CFPE approximation is valid provided two conditions are satisfied. A large number of molecules suffices to ensure that both conditions are satisfied however this is NOT a necessary condition. This suggests that there are regimes in which the particle numbers may not be very large and yet the CFPE may still provide a reasonably good approximation of the CME. However Gillespie's derivation does not provide us with a means to estimate the accuracy of the CFPE for general chemical systems.

Questions regarding the validity and accuracy of the CFPE and CLE are more important now than ever before. In the past decade, interest has virtually exploded in realistic stochastic simulations of biochemical reactions inside cells \cite{Turner2004,GrimaSchnell,Paulsson2005,Meng2004}. The exact method of sampling the trajectories of the CME, the stochastic simulation algorithm \cite{Gillespie1977}, is computationally expensive and the CME is analytically intractable; thus approximate methods such as the CFPE and the CLE have come to the foreground as an alternative means to obtain numerical and theoretical insight into the functioning of intracellular biochemical networks \cite{Hou2003,Simpson2004,Xiao2007,Wilkie2008,Sotiropoulos2009}. These networks are typically characterized by a large number of bimolecular reactions in which at least one of the species is present in very small molecule numbers \cite{GrimaSchnell,Ishihama2008,Ghaemmaghami2003}, indeed the precise conditions in which the fidelity of the CFPE remains unclear. Hence the question of the accuracy of the CFPE has nowadays become a practical one -- how much can we trust the conclusions derived from the CFPE or the corresponding CLE? 

In this article, we derive formulas to estimate the relative error in the CFPE predictions of the mean concentrations and of the variance of the fluctuations about the mean. The results are valid for all monostable chemical reaction networks. As a byproduct of our derivation, we will also clarify the connection between the CFPE and van Kampen's system-size expansion, in particular showing that the non-linear character of the CFPE is not completely spurious and that generally CFPE estimates are more accurate than those obtained from the linear Fokker-Planck equation. The article is organized as follows. In Section II, we use the multivariate system-size expansion to derive expressions for the mean concentrations and for the variance of the fluctuations as predicted by the CME accurate to order $O(\Omega^{-2})$. In Section III, we develop the system-size expansion of the CFPE and use it to derive expressions for the mean concentrations and for the variance of the fluctuations accurate to the same order as derived for the CME in Section II. In Section IV, we use the results of the previous two sections to derive expressions for the relative error in the predictions of the CFPE. We also compare the predictions of the CFPE and the linear Fokker-Planck equation. These results are tested on two bimolecular reaction systems -- dimerization and an enzyme-catalyzed reaction -- in Section V.  We conclude by a discussion in Section VI.

\section{Perturbative expansion of the CME}

\subsection{The Multivariate System-Size Expansion of the CME} 

We will now probe the mesoscopic description provided by the CME using the system-size expansion developed by van Kampen \cite{vanKampen}. This method allows one to derive expressions for the mean concentrations and for the variance of the fluctuations about these concentrations, as predicted by the CME, accurate to the order of any desired power of the inverse square root of the volume. The only requirement for the expansion to hold is that the steady-state of the chemical system is asymptotically stable. For the applications that we are interested in, namely biochemical reactions in intracellular conditions, the number of molecules can be very small, in some cases just few tens of molecules of a given species per cell. We will derive equations accurate to $O(\Omega^{-2})$ -- this accuracy should be more than sufficient for the applications mentioned since terms of lower order, $O(\Omega^{-1})$, already imply corrections to the concentrations of the order of a single molecule in the compartment. To our knowledge this is the first time that the system-size expansion has been carried to this order for a general system of $N$ interacting chemical species. van Kampen has treated a one species example to the same order in his book \cite{vanKampen} while Elf and Ehrenberg \cite{ElfEhrenberg} have derived the multivariate expansion to $O(\Omega^0)$.    

The starting point of the system-size expansion is to write the absolute number of molecules of species $i$ as:
\begin{equation}
\frac{n_i}{\Omega} = \phi_i + \Omega^{-1/2} \epsilon_i,
\end{equation}
where $\phi_i$ is the macroscopic concentration of species $i$ as determined by the REs. This has the effect of transforming all functions of $n_i$ in the CME into functions of $\epsilon_i$. The expansion of the CME proceeds by writing Eq. (2) in terms of the new variables. Details of this transformation can be found in \cite{Grima2010}; here we will simply state the relevant results and use them for our present derivation. The variable change causes the probability distribution of molecular populations, $P(\vec{n},t)$, to be transformed into the probability distribution of fluctuations, $\Pi(\vec{\epsilon},t)$, where $\vec{\epsilon}=(\epsilon_1,...,\epsilon_N)^T$. The time derivative, the step operator and the microscopic rate function in the CME, read in the new variables:
\begin{align}
\frac{\partial P(\vec{n},t)}{\partial t} &= \frac{\partial \Pi(\vec{\epsilon},t)}{\partial t} - \Omega^{1/2} \sum_{i=1}^N \frac{\partial \phi_i}{\partial t} \frac{\partial \Pi(\vec{\epsilon},t)}{\partial \epsilon_i}, \\
\displaystyle\prod_{i=1}^N E_i^{-S_{ij}} - 1 &= \sum_{k=1}^{\infty} -1^{k} \Omega^{-k/2} a_j^k, \\
\hat{f}_j &= \sum_{k=0}^{2} \Omega^{-k/2} b_j^k + c_j^2 \Omega^{-1} + c_j^3 \Omega^{-3/2},
\end{align}
where
\begin{align}
a_j^k &= \frac{1}{k!} \biggl( \sum_{i=1}^N S_{ij} \frac{\partial}{\partial \epsilon_i} \biggr)^k, \\
b_j^k &= \frac{1}{k!} \biggl( \sum_{w=1}^N \epsilon_w \frac{\partial}{\partial \phi_w} \biggr)^k f_j(\vec{\phi}), \\
c_j^2 &= -\frac{1}{2} \sum_{w=1}^N \phi_w \frac{\partial^2 f_j(\vec{\phi})}{\partial \phi_w^2}, \\
c_j^3 &= -\frac{1}{2} \sum_{w=1}^N \epsilon_w \frac{\partial^2 f_j(\vec{\phi})}{\partial \phi_w^2}.
\end{align}
Note that in Eq. (9) the microscopic rate function is expressed in terms of the macroscopic rate function. As we shall shortly see, this is convenient from a calculation point of view since the final expressions for the means and variances will be solely in terms of functions which appear in the REs. Note that the upper limit of the sum in Eq. (9) is 2 because all reactions involve at most the interaction of two molecules and hence $b_j^k$ equals zero for $k > 2$. Although our analysis is specifically for elementary reactions, one can easily extend the approach to include  
``elementary complex'' reactions  \cite{ElfEhrenberg}. However we shall not pursue this here.

Substituting Eqs. (7-9) in Eq. (2) we get the following new form of the CME:
\begin{align}
\frac{\partial \Pi(\vec{\epsilon},t)}{\partial t} = & \Omega^0 \sum_{j=1}^R (a_j^2 b_j^0 - a_j^1 b_j^1) \Pi(\vec{\epsilon},t) + \nonumber \\ & \Omega^{-1/2}  \sum_{j=1}^R (a_j^2 b_j^1 - a_j^1 b_j^2 - a_j^1 c_j^2 - a_j^3 b_j^0) \Pi(\vec{\epsilon},t) + \nonumber \\ & \Omega^{-1}  \sum_{j=1}^R (a_j^2 b_j^2 + a_j^2 c_j^2 + a_j^4 b_j^0 - a_j^1 c_j^3 - a_j^3 b_j^1) \Pi(\vec{\epsilon},t) + \nonumber \\ & \Omega^{-3/2}  \sum_{j=1}^R (a_j^2 c_j^3 + a_j^4 b_j^1 - a_j^3 b_j^2 - a_j^3 c_j^2 - a_j^5 b_j^0) \Pi(\vec{\epsilon},t) + O(\Omega^{-2}).
\end{align}
Note that terms proportional to $\Omega^{1/2}$ do not appear in the expansion of the CME. This is because when one substitutes Eqs. (7-9) in Eq. (2), one equates terms of this order on both sides of the CME which simply gives us back the macroscopic REs, Eq. (3).

To proceed further we need the explicit dependence of the right hand side of Eq. (14) on the new variables $\epsilon_i$. This is obtained by substituting Eqs. (10-13) in Eq. (14) which leads to:
\begin{align}
\frac{\partial \Pi(\vec{\epsilon},t)}{\partial t} = & \Omega^0 \biggl(-J_i^w \partial_i (\epsilon_w \Pi) + \frac{1}{2} D_{ip} \partial_{ip}^2 \Pi \biggr) + \nonumber \\ &\Omega^{-1/2} \biggl( -\frac{1}{2} J_i^{wp} \partial_i (\epsilon_w \epsilon_p \Pi) + \frac{1}{2} \phi_w J_i^{w(2)} \partial_i \Pi + \frac{1}{2} J_{ip}^w \partial_{ip}^2 (\epsilon_w \Pi) - \frac{1}{6} D_{ipw} \partial_{ipw}^3 \Pi  \biggr) + \nonumber \\ & \Omega^{-1} \biggr( \frac{1}{2} J_i^{w(2)} \partial_i (\epsilon_w \Pi) + \frac{1}{4} J_{ip}^{wm} \partial_{ip}^2 (\epsilon_w \epsilon_m \Pi) - \frac{1}{4} J_{ip}^{w(2)} \phi_w \partial_{ip}^2 \Pi  -\frac{1}{6} J_{ipm}^w \partial_{ipm}^3 (\epsilon_w \Pi) + \nonumber \\  &+ \frac{1}{24} D_{ipmw} \partial_{ipmw}^4 \Pi \biggl) + \Omega^{-3/2} \biggl( -\frac{1}{4} J_{ip}^{w(2)} \partial_{ip}(\epsilon_w \Pi) + \frac{1}{24} J_{ipmr}^w \partial_{ipmr} (\epsilon_w \Pi) - \frac{1}{12} J_{ipm}^{wk} \nonumber \\ & \times \partial_{ipm} (\epsilon_w \epsilon_k \Pi) + \frac{1}{12} J_{ipm}^{w(2)} \phi_w \partial_{ipm} \Pi - \frac{1}{120} D_{ipmrs} \partial_{ipmrs} \Pi \biggr) + O(\Omega^{-2}).
\end{align}  
Note that in the above equation, we have used the Einstein summation convention where all twice repeated indices are understood to be summed over 1 to $N$. The partial derivative $\partial_{i..j}^n$ denotes $\partial^n / \partial \epsilon_i .. \partial \epsilon_j$. We have also used the following two convenient definitions:
\begin{align}
D_{ij..r} &= \sum_{k=1}^R S_{ik} S_{jk} ... S_{rk} f_k(\vec{\phi}), \\
J_{ij..r}^{st..z} &= \frac{\partial}{\partial \phi_s}  \frac{\partial}{\partial \phi_t} ... \frac{\partial}{\partial \phi_z} D_{ij..r}, \quad J_{ij..r}^{s(2)} = J_{ij..r}^{ss}.
\end{align}
From Eq. (3) it follows that $D_i = \partial \phi_i /\partial t$ and consequently $J_{i}^{s}$ represents the $i$-$s$ element of the Jacobian matrix associated with the REs of the system. 

Note that Eq. (15) to order $\Omega^0$ is the linear Fokker-Planck equation which was mentioned in the introduction. The drift vector is linear in the $\epsilon$ variables while the diffusion tensor is independent of them. Both depend on time via their own dependence on the macroscopic concentrations. This level of approximation is frequently called the linear-noise approximation, a nowadays popular means of estimating the size of the concentration fluctuations about the macroscopic concentrations \cite{ElfEhrenberg}. We are interested in the dynamics on mesoscopic length scales and hence we shall consider terms of higher order than $\Omega^0$ in Eq. (15).

\subsection{Time-evolution equations for the moments}

We now proceed to construct equations for the moments of the $\epsilon$ variables. We start by expanding $\Pi(\vec{\epsilon},t)$ as a series in powers of the inverse square root of the volume:
\begin{equation}
\Pi(\vec{\epsilon},t) = \sum_{j=0}^{\infty }\Pi_j(\vec{\epsilon},t) \Omega^{-j/2},
\end{equation}
from which it follows that the moments possess an equivalent expansion:
\begin{equation}
\langle \epsilon_k \epsilon_m ... \epsilon_r \rangle =  \sum_{j=0}^{\infty } [ \epsilon_k \epsilon_m ... \epsilon_r ]_j \Omega^{-j/2},
\end{equation}
where
\begin{equation}
[ \epsilon_k \epsilon_m ... \epsilon_r ]_j = \int \epsilon_k \epsilon_m ... \epsilon_r \ \Pi_j(\vec{\epsilon},t) d\vec{\epsilon}.
\end{equation}
The angled brackets denote the statistical average. Some subtle points associated with the perturbative expansion in the probability density and with the physical interpretation of $[ \epsilon_k \epsilon_m ... \epsilon_r ]_j$ are discussed in Appendix A. The time-evolution equations for the moments are obtained as follows. One starts by substituting Eq. (18) in Eq. (15), multiplying the resulting equation on both sides by $ \epsilon_k \epsilon_m ... \epsilon_r$ and integrating over $d\vec{\epsilon}$. Equating terms of order $\Omega^{-j/2}$ on both sides of the equation gives the time-evolution equation for $[ \epsilon_k \epsilon_m ... \epsilon_r ]_j$.  Finally one constructs the time-evolution equation for the moments using Eq. (19).

As mentioned earlier, our aim is to determine the mean concentrations and the variance of the fluctuations about the means and hence we must relate the latter to the moments of the $\epsilon$ variables above. Using Eqs. (6) and (19), one can easily verify that the mean concentration of species $i$ and the variance of the fluctuations about it, accurate to order $\Omega^{-2}$ are respectively given by:
\begin{align}
\biggl \langle \frac{n_i}{\Omega} \biggr \rangle &= \phi_i + \Omega^{-1/2} \langle \epsilon_i \rangle = \phi_i + \Omega^{-1/2} \sum_{j=0}^{3} [ \epsilon_i ]_j \Omega^{-j/2} + O(\Omega^{-5/2}), \\
\sigma_i^2 &= \biggl \langle \biggl(\frac{n_i}{\Omega} \biggr)^2 \biggr \rangle - \biggl \langle \frac{n_i}{\Omega} \biggr \rangle^2 = \Omega^{-1} (\langle \epsilon_i^2 \rangle - \langle \epsilon_i \rangle^2) \nonumber \\ &= \Omega^{-1} \biggr( \sum_{j=0}^{2} [ \epsilon_i^2 ]_j \Omega^{-j/2} - \biggl(\sum_{j=0}^{1} [ \epsilon_i ]_j \Omega^{-j/2} \biggr)^2 - \Omega^{-1}  [ \epsilon_i ]_0  [ \epsilon_i ]_2 \biggl) + O(\Omega^{-5/2}).
\end{align}
Hence it is clear that to determine the mean and variance accurate to order $\Omega^{-2}$, we shall need to determine the first and second moments of the $\epsilon$ variables accurate to orders $\Omega^{-3/2}$ and $\Omega^{-1}$ respectively. 

We proceed by implementing the calculation recipe outlined just after Eq. (20) to derive equations for the corrections to the second moments accurate to order $\Omega^{-1}$:
\begin{align}
\frac{\partial}{\partial t} [ \epsilon_r \epsilon_k ]_0 &=  J_r^{w}  [ \epsilon_w \epsilon_k ]_0 + (r \leftrightarrow k) + D_{rk}, \\
\frac{\partial}{\partial t} [ \epsilon_r \epsilon_k ]_1 &=  J_r^{w}  [ \epsilon_w \epsilon_k ]_1 + \frac{1}{2} J_r^{wp}  [ \epsilon_w \epsilon_p \epsilon_k ]_0 - \frac{1}{2} J_r^{w(2)}  \phi_w [ \epsilon_k ]_0 \nonumber \\ &  + (r \leftrightarrow k)  + J_{kr}^w [ \epsilon_w ]_0, \\
\frac{\partial}{\partial t} [ \epsilon_r \epsilon_k ]_2 &=  J_r^{w}  [ \epsilon_w \epsilon_k ]_2 + \frac{1}{2} J_r^{wp}  [ \epsilon_w \epsilon_p \epsilon_k ]_1 - \frac{1}{2} J_r^{w(2)}  \phi_w [ \epsilon_k ]_1 \nonumber \\ & - \frac{1}{2}  J_r^{w(2)}  [ \epsilon_w \epsilon_k ]_0 + (r \leftrightarrow k)  +  J_{kr}^w [ \epsilon_w ]_1 + \frac{1}{2} J_{rk}^{wm} [ \epsilon_w \epsilon_m ]_0 - \frac{1}{2} J_{rk}^{w(2)} \phi_w.
\end{align}
Details of the calculations leading to the above equations are illustrated by a step-by-step derivation of Eq. (25) in Appendix B. Note that the short-hand notation $(r \leftrightarrow k)$  stands for all the expressions of the same form as the ones preceding the notation but with $r$ and $k$ interchanged. For example in Eq. (24), $(r \leftrightarrow k)$ stands for $J_k^{w}  [ \epsilon_w \epsilon_r ]_1 + \frac{1}{2} J_k^{wp}  [ \epsilon_w \epsilon_p \epsilon_r ]_0 - \frac{1}{2} J_k^{w(2)}  \phi_w [ \epsilon_r ]_0$. This notation will be used throughout the rest of the article since it enables the equations to be written in a compact way.

The equation for $[ \epsilon_r \epsilon_k ]_0$, Eq. (23), is a Lyapunov equation which can be solved either analytically (see for example \cite{ElfEhrenberg,Keizer}) or else numerically, for example using the built in functions of Matlab and Mathematica. Solution of the equation for $[ \epsilon_r \epsilon_k ]_1$, Eq. (24), requires the solutions of the equations for the first and third moments to order $\Omega^0$:
\begin{align}
\frac{\partial}{\partial t} [ \epsilon_r ]_0 &=  J_r^{w}  [ \epsilon_w ]_0, \\
\frac{\partial}{\partial t} [ \epsilon_r \epsilon_k \epsilon_l ]_0 &=  J_l^{w}  [ \epsilon_w \epsilon_k \epsilon_r ]_0 + (l \leftrightarrow k) + (k \leftrightarrow r) \nonumber \\ &+ D_{rl} [ \epsilon_k ]_0 +  (k \leftrightarrow l) + (r \leftrightarrow l).
\end{align}
Note that in Eq. (27), $(l \leftrightarrow k) + (k \leftrightarrow r)$ stands for two expressions; the first expression corresponds to the first term on the right hand side of Eq. (27) with $l$ and $k$ interchanged and the second expression is the first expression just obtained with $k$ and $r$ interchanged. By a similar reasoning, it follows that $(k \leftrightarrow l) + (r \leftrightarrow l)$ in Eq. (27) stands for $D_{rk} [ \epsilon_l ]_0 + D_{lk} [ \epsilon_r ]_0$. Note that in steady-state conditions, $[ \epsilon_r ]_0 = [ \epsilon_r \epsilon_k \epsilon_l ]_0 = 0$ and and consequently there is no correction to the second moments to $O(\Omega^{-1})$, i.e., $[ \epsilon_r \epsilon_k ]_1 = 0$. 

Solution of the equation for $[ \epsilon_r \epsilon_k ]_2$, Eq. (25), requires the solutions of the corrections to the the first and third moments to order $\Omega^{-1/2}$ and the second and fourth moments to order $\Omega^{0}$ :
\begin{align}
\frac{\partial}{\partial t} [ \epsilon_r ]_1 &=  J_r^{w}  [ \epsilon_w ]_1 + \frac{1}{2} J_r^{wp}  [ \epsilon_w \epsilon_p ]_0 - \frac{1}{2} J_r^{w(2)}  \phi_w, \\
\frac{\partial}{\partial t} [ \epsilon_r \epsilon_k \epsilon_l ]_1 &=  J_l^{w}  [ \epsilon_w \epsilon_k \epsilon_r ]_1 + \frac{1}{2} J_l^{wp} [ \epsilon_w \epsilon_p \epsilon_r \epsilon_k ]_0 - \frac{1}{2} J_l^{w(2)} \phi_w [ \epsilon_r \epsilon_k ]_0 \nonumber \\ &+ (l \leftrightarrow k) + (k \leftrightarrow r) + D_{rl} [ \epsilon_k ]_1 +  J_{rl}^w [ \epsilon_w \epsilon_k ]_0 + (k \leftrightarrow l) \nonumber \\ & + (r \leftrightarrow l) +  D_{rkl}, \\
\frac{\partial}{\partial t} [ \epsilon_r \epsilon_k \epsilon_l \epsilon_m ]_0 &=  J_r^{w}  [ \epsilon_w \epsilon_k \epsilon_l \epsilon_m ]_0 + (r \leftrightarrow m) + (m \leftrightarrow k) + (k \leftrightarrow l) + D_{rm} [ \epsilon_k \epsilon_l ]_0 \nonumber \\ & +  (m \leftrightarrow l) + (l \leftrightarrow k) + (r \leftrightarrow m) + (m \leftrightarrow l) + (k \leftrightarrow m).
\end{align}
The procedure to obtain the second moments to order $\Omega^{-1}$ is now clear. One first solves Eq. (23) to get $[ \epsilon_r \epsilon_k ]_0$; then one solves Eqs. (26-27) and substitutes in Eq. (24) to get $[ \epsilon_r \epsilon_k ]_1$; finally one solves Eqs. (28-30) and substitutes these, together with the solution of Eq. (23), in Eq. (25) to get $[ \epsilon_r \epsilon_k ]_2$.

The first moment equations and the corresponding equations for the mean concentrations can be obtained in an analogous manner as for the second moments. The equations for $[ \epsilon_r ]_0$ and $[ \epsilon_r ]_1$ have been already derived, Eqs. (26) and (28), respectively. The equations for $[ \epsilon_r ]_2$ and $[ \epsilon_r ]_3$ are given by:
\begin{align}
\frac{\partial}{\partial t} [ \epsilon_r ]_2 &=  J_r^{w}  [ \epsilon_w ]_2 + \frac{1}{2} J_r^{wp}  [ \epsilon_w \epsilon_p ]_1 - \frac{1}{2} J_r^{w(2)} [ \epsilon_w ]_0, \\
\frac{\partial}{\partial t} [ \epsilon_r ]_3 &=  J_r^{w}  [ \epsilon_w ]_3 + \frac{1}{2} J_r^{wp}  [ \epsilon_w \epsilon_p ]_2 - \frac{1}{2} J_r^{w(2)} [ \epsilon_w ]_1.
\end{align} 
The procedure to obtain the first moments to order $\Omega^{-3/2}$ is now also clear. One first solves Eq. (26) to get $[ \epsilon_r ]_0$; then one solves Eq. (23) and substitutes in Eq. (28) to obtain $[ \epsilon_r ]_1$; finally one uses the solutions already obtained when deriving the second moments to solve Eqs. (31-32) for $[ \epsilon_r ]_2$ and $[ \epsilon_r ]_3$. 

Given the first and second moments accurate to $\Omega^{-3/2}$ and $\Omega^{-1}$, one finally determines the mean concentrations and the variance of the fluctuations about them accurate to $\Omega^{-2}$ from Eqs. (21-22). Although the procedure of obtaining the latter final expressions is fairly laborious, as we shall see in the next section, in order to obtain the leading order error in the predictions of the CFPE,  it will only be necessary for us to solve very few of these equations explicitly.  

\section{Perturbative expansion of the CFPE}

In this section we use the system-size expansion to derive expressions for the mean concentrations and the fluctuations about them, as predicted by the CFPE, accurate to $O(\Omega^{-2})$. To the best of our knowledge this is the first time that the expansion has been used on the CFPE although the method is similar in principle to the small-noise expansion of Fokker-Planck equations as presented by Gardiner \cite{Gardiner}. The CFPE is obtained by truncating the Kramer's Moyal expansion to include at most second-order derivatives:
\begin{align}
\frac{\partial P(\vec{n},t)}{\partial t} &= \Omega \sum_{j=1}^{R} \biggl( \displaystyle\prod_{i=1}^N E_i^{-S_{ij}} - 1 \biggr) \hat{f}_j(\vec{n},\Omega) P(\vec{n},t) \nonumber \\ &= \Omega \sum_{j=1}^{R} \biggl( \displaystyle\prod_{i=1}^N e^{-S_{ij} \partial/\partial n_i} - 1 \biggr) \hat{f}_j(\vec{n},\Omega) P(\vec{n},t) \nonumber \\ &\simeq \Omega \sum_{j=1}^{R} \biggl( -\sum_{i=1}^N S_{ij} \frac{\partial}{\partial n_i} + \frac{1}{2} \sum_{i,w=1}^N S_{ij} S_{wj} \frac{\partial^2}{\partial n_i \partial n_w} \biggr) \hat{f}_j(\vec{n},\Omega) P(\vec{n},t).
\end{align}
Note that the second step above, follows by Taylor expanding the step operator.

Next we perform the system-size expansion on the CFPE, Eq. (33), i.e., we make the variable transformation given by Eq. (6) which transforms functions of $n_i$ into functions of the new variables $\epsilon_i$. The probability distribution $P(\vec{n},t)$ is transformed into a new one $\Pi_{F}(\vec{\epsilon},t)$. Note that the subscript $F$ will be used to distinguish quantities calculated using the CFPE from those previously calculated using the CME. The time derivative on the left hand side of the equation and the microscopic rate function $\hat{f}_j(\vec{n})$ transform as in the case of the CME and are given by Eqs. (7) and (9) together with the definitions Eqs. (11-13) and with $\Pi(\vec{\epsilon},t)$ replaced by $\Pi_{F}(\vec{\epsilon},t)$. The operators involving derivatives with respect to absolute particle number transform as follows:
\begin{align}
\sum_{i=1}^N S_{ij} \frac{\partial}{\partial n_i} &= \Omega^{-1/2} a_j^1, \\
\frac{1}{2} \sum_{i,w=1}^N S_{ij} S_{wj} \frac{\partial^2}{\partial n_i \partial n_w} &= \Omega^{-1} a_j^2,
\end{align}
where the operators $a_j^k$ are as defined in Eq. (10). Hence the CFPE in the new variables reads:
\begin{align}
\frac{\partial \Pi_F(\vec{\epsilon},t)}{\partial t} = & \Omega^0 \sum_{j=1}^R (a_j^2 b_j^0 - a_j^1 b_j^1) \Pi_F(\vec{\epsilon},t) + \nonumber \\ & \Omega^{-1/2}  \sum_{j=1}^R (a_j^2 b_j^1 - a_j^1 b_j^2 - a_j^1 c_j^2) \Pi_F(\vec{\epsilon},t) + \nonumber \\ & \Omega^{-1}  \sum_{j=1}^R (a_j^2 b_j^2 + a_j^2 c_j^2  - a_j^1 c_j^3) \Pi_F(\vec{\epsilon},t) + \Omega^{-3/2}  \sum_{j=1}^R a_j^2 c_j^3 \Pi_F(\vec{\epsilon},t).
\end{align}
Note that whereas the transformation given by Eq. (6) on the CME leads to an infinite series in powers of the inverse square root of the volume, Eq. (14), the same transformation on the CFPE leads to a finite series with the highest order term being of order $\Omega^{-3/2}$ (this is  only true for elementary reactions). 

The derivation of the equations for the time evolution of the moments of the $\epsilon$ variables proceeds in an exactly analogous manner as to that presented in detail in section II. The probability distribution is written as a series in powers of the inverse square root of the volume,
\begin{equation}
\Pi_F(\vec{\epsilon},t) = \sum_{j=0}^{3 }\Pi_{F,j}(\vec{\epsilon},t) \Omega^{-j/2}.
\end{equation}
and the moments are then generally given by:
\begin{equation}
\langle \epsilon_k \epsilon_m ... \epsilon_r \rangle_F =  \sum_{j=0}^{3} [ \epsilon_k \epsilon_m ... \epsilon_r ]_{F,j} \Omega^{-j/2},
\end{equation}
where
\begin{equation}
[ \epsilon_k \epsilon_m ... \epsilon_r ]_{F,j} = \int \epsilon_k \epsilon_m ... \epsilon_r \ \Pi_{F,j}(\vec{\epsilon},t) d\vec{\epsilon}.
\end{equation}
The equations for the mean concentrations and the variance of the fluctuations about them are given by Eqs. (21-22) with the subscript $F$ carried throughout.  The time evolution equations for the corrections to the moments can be derived as before. Although there is some repetition involved, we will state these equations in full so that the differences between them and those derived using the CME are very clear.

The equations for the corrections to the second moments accurate to order $\Omega^{-1}$ are:
\begin{align}
\frac{\partial}{\partial t} [ \epsilon_r \epsilon_k ]_{F,0} &=  J_r^{w}  [ \epsilon_w \epsilon_k ]_{F,0} + (r \leftrightarrow k) + D_{rk}, \\
\frac{\partial}{\partial t} [ \epsilon_r \epsilon_k ]_{F,1} &=  J_r^{w}  [ \epsilon_w \epsilon_k ]_{F,1} + \frac{1}{2} J_r^{wp}  [ \epsilon_w \epsilon_p \epsilon_k ]_{F,0} - \frac{1}{2} J_r^{w(2)}  \phi_w [ \epsilon_k ]_{F,0} \nonumber \\ &  + (r \leftrightarrow k)  + J_{kr}^w [ \epsilon_w ]_{F,0}, \\
\frac{\partial}{\partial t} [ \epsilon_r \epsilon_k ]_{F,2} &=  J_r^{w}  [ \epsilon_w \epsilon_k ]_{F,2} + \frac{1}{2} J_r^{wp}  [ \epsilon_w \epsilon_p \epsilon_k ]_{F,1} - \frac{1}{2} J_r^{w(2)}  \phi_w [ \epsilon_k ]_{F,1} \nonumber \\ & - \frac{1}{2}  J_r^{w(2)}  [ \epsilon_w \epsilon_k ]_{F,0} + (r \leftrightarrow k)  +  J_{kr}^w [ \epsilon_w ]_{F,1} + \frac{1}{2} J_{rk}^{wm} [ \epsilon_w \epsilon_m ]_{F,0} - \frac{1}{2} J_{rk}^{w(2)} \phi_w.
\end{align}
Note that these are the same as Eqs (23-25) but with subscript $F$; the implicit reason for this is that only terms containing $a_j^1$ and $a_j^2$ contribute to the equations for the second moments and all such terms are equally present in Eqs. (14) and (36). Note also that Eqs. (23) and (40) lead to the same solution, i.e., $[\epsilon_r \epsilon_k ]_{0} =[\epsilon_r \epsilon_k ]_{F,0}$.  The solution of $[\epsilon_r \epsilon_k ]_{F,1}$ is dependent on the solutions of the time evolution equations for $[\epsilon_r]_{F,0}$ and $[\epsilon_r \epsilon_k \epsilon_l]_{F,0}$. The equations for the latter are the same as Eqs. (26-27) but with subscript $F$; this is since Eq. (14) and (36) are equal to order $\Omega^0$. It follows that $[\epsilon_r ]_{0} =[\epsilon_r ]_{F,0}$ and $[\epsilon_r \epsilon_k \epsilon_l]_{0} =[\epsilon_r \epsilon_k \epsilon_l]_{F,0}$ from which we can conclude using Eq. (41) that $[\epsilon_r \epsilon_k ]_{1} =[\epsilon_r \epsilon_k ]_{F,1}$. However, as we now show, generally $[\epsilon_r \epsilon_k ]_{2} \ne [\epsilon_r \epsilon_k ]_{F,2}$.

The solution of $[\epsilon_r \epsilon_k ]_{F,2}$ is dependent on the solutions of the time evolution equations for $[\epsilon_r]_{F,1}$, $[\epsilon_r \epsilon_k \epsilon_l]_{F,1}$ and $[\epsilon_r \epsilon_k \epsilon_l \epsilon_m]_{F,0}$ which are:
\begin{align}
\frac{\partial}{\partial t} [ \epsilon_r ]_{F,1} &=  J_r^{w}  [ \epsilon_w ]_{F,1} + \frac{1}{2} J_r^{wp}  [ \epsilon_w \epsilon_p ]_{F,0} - \frac{1}{2} J_r^{w(2)}  \phi_w, \\
\frac{\partial}{\partial t} [ \epsilon_r \epsilon_k \epsilon_l ]_{F,1} &=  J_l^{w}  [ \epsilon_w \epsilon_k \epsilon_r ]_{F,1} + \frac{1}{2} J_l^{wp} [ \epsilon_w \epsilon_p \epsilon_r \epsilon_k ]_{F,0} - \frac{1}{2} J_l^{w(2)} \phi_w [ \epsilon_r \epsilon_k ]_{F,0} \nonumber \\ &+ (l \leftrightarrow k) + (k \leftrightarrow r) + D_{rl} [ \epsilon_k ]_{F,1} +  J_{rl}^w [ \epsilon_w \epsilon_k ]_{F,0} + (k \leftrightarrow l) \nonumber \\ & + (r \leftrightarrow l), \\
\frac{\partial}{\partial t} [ \epsilon_r \epsilon_k \epsilon_l \epsilon_m ]_{F,0} &=  J_r^{w}  [ \epsilon_w \epsilon_k \epsilon_l \epsilon_m ]_{F,0} + (r \leftrightarrow m) + (m \leftrightarrow k) + (k \leftrightarrow l) + D_{rm} [ \epsilon_k \epsilon_l ]_{F,0} \nonumber \\ & +  (m \leftrightarrow l) + (l \leftrightarrow k) + (r \leftrightarrow m) + (m \leftrightarrow l) + (k \leftrightarrow m).
\end{align}
Equations (43) and (45) have the same form as Eqs. (28) and (30) respectively. This combined with the fact that the right hand sides of Eqs. (43) and (45) are functions of $[\epsilon_r \epsilon_k ]_{F,0}$ and that $[\epsilon_r \epsilon_k ]_{0} =[\epsilon_r \epsilon_k ]_{F,0}$, implies that $[\epsilon_r ]_{1} =[\epsilon_r]_{F,1}$ and $[\epsilon_r \epsilon_k \epsilon_l \epsilon_m ]_{0} =[\epsilon_r \epsilon_k \epsilon_l \epsilon_m]_{F,0}$. However note that Eq. (44) has one term missing compared to its counterpart Eq. (29) and hence generally $[\epsilon_r \epsilon_k \epsilon_l]_{1} \ne [\epsilon_r \epsilon_k \epsilon_l ]_{F,1}$ from which it follows using Eq. (42) that $[\epsilon_r \epsilon_k ]_{2} \ne [\epsilon_r \epsilon_k ]_{F,2}$. 

The only remaining equations to be considered are those paralleling Eqs. (31) and (32) for which we find:
\begin{align}
\frac{\partial}{\partial t} [ \epsilon_r ]_{F,2} &=  J_r^{w}  [ \epsilon_w ]_{F,2} + \frac{1}{2} J_r^{wp}  [ \epsilon_w \epsilon_p ]_{F,1} - \frac{1}{2} J_r^{w(2)} [ \epsilon_w ]_{F,0}, \\
\frac{\partial}{\partial t} [ \epsilon_r ]_{F,3} &=  J_r^{w}  [ \epsilon_w ]_{F,3} + \frac{1}{2} J_r^{wp}  [ \epsilon_w \epsilon_p ]_{F,2} - \frac{1}{2} J_r^{w(2)} [ \epsilon_w ]_{F,1}.
\end{align} 
By similar arguments to the above, these equations imply $[\epsilon_r ]_{2} =[\epsilon_r]_{F,2}$ and $[\epsilon_r ]_{3}  \ne [\epsilon_r]_{F,3}$.

Hence, in summary, we have obtained the following results:
\begin{enumerate}
\item $[\epsilon_r]_{0} =[\epsilon_r]_{F,0}$, $[\epsilon_r \epsilon_k ]_{0} =[\epsilon_r \epsilon_k ]_{F,0}$ , $[\epsilon_r \epsilon_k \epsilon_l]_{0} =[\epsilon_r \epsilon_k \epsilon_l]_{F,0}$, $[\epsilon_r \epsilon_k \epsilon_l \epsilon_m ]_{0} =[\epsilon_r \epsilon_k \epsilon_l \epsilon_m]_{F,0}$
\item $[\epsilon_r ]_{1} =[\epsilon_r]_{F,1}$, $[\epsilon_r \epsilon_k ]_{1} =[\epsilon_r \epsilon_k ]_{F,1}$, $[\epsilon_r \epsilon_k \epsilon_l]_{1} \ne [\epsilon_r \epsilon_k \epsilon_l ]_{F,1}$
\item $[\epsilon_r ]_{2} =[\epsilon_r]_{F,2}$, $[\epsilon_r \epsilon_k ]_{2} \ne [\epsilon_r \epsilon_k ]_{F,2}$
\item $[\epsilon_r ]_{3}  \ne [\epsilon_r]_{F,3}$
\end{enumerate}
Note that these results are not for the moments but for the corrections to the moments; the real physical meaning of these results in terms of means and variances will be elucidated in the next section.

Using Eqs. (32) and (47), Eqs (25) and (42) and Eqs. (29) and (44), we can respectively write down simple equations for the differences in the corrections to the first, second and third moments as predicted by the CFPE and the CME:
\begin{align}
\frac{\partial}{\partial t} \Delta_r &=  J_r^{w}  \Delta_w  + \frac{1}{2} J_r^{wp}  \Delta_{wp}, \\
\frac{\partial}{\partial t} \Delta_{rk}  &=  J_r^{w}  \Delta_{wk}  + J_k^{w}  \Delta_{wr}  + \frac{1}{2}( J_r^{wp}  \Delta_{wpk} + J_k^{wp}  \Delta_{wpr} ), \\
\frac{\partial}{\partial t} \Delta_{rkl}  &=  J_l^{w}  \Delta_{wkr}  + J_k^{w}  \Delta_{wlr}  + J_r^{w}  \Delta_{wlk} +D_{rkl},
\end{align}
where we have used the convenient definitions:
\begin{align}
\Delta_r &= [\epsilon_r ]_{3}  - [\epsilon_r]_{F,3}, \\
\Delta_{rk}  &= [\epsilon_r \epsilon_k ]_{2} - [\epsilon_r \epsilon_k ]_{F,2}, \\
\Delta_{rkl} &= [\epsilon_r \epsilon_k \epsilon_l]_{1} - [\epsilon_r \epsilon_k \epsilon_l ]_{F,1}.
\end{align}

\section{Comparison of the predictions of the CFPE and the CME}

In this section we will use the results derived in the last section to obtain formulas for the absolute and relative errors (to leading order) in the CFPE predictions of the mean concentrations and the variance of the fluctuations. Using these formulas we will be able to deduce the general conditions in which the differences between the CFPE and the CME are minimal. Furthermore we will show that the CFPE is generally more accurate than the linear Fokker-Planck equation of van Kampen and that the mean concentrations of the CFPE to order $\Omega^{-1}$ are precisely the same as those obtained from Effective Mesoscopic Rate Equations.

\subsection{Estimating the absolute and relative errors in CFPE predictions}

We will now derive expressions for the leading order term of the absolute and relative errors made by the CFPE in predicting the mean concentrations and the variance of the fluctuations about the mean concentrations. We will also obtain an expression for the leading order term of the absolute error made by the CFPE in predicting the skewness of the probability distribution of the concentrations.

The mean concentration predicted by the CME, $\langle n_i / \Omega \rangle$, is given by Eq. (21) while the mean concentration predicted by the CFPE, $\langle n_i / \Omega \rangle_F$ is given by the same equation but with the subscript $F$ carried throughout. Subtracting the two expressions and using the summary of results in Section III together with Eq. (51) we get the absolute error in the CFPE concentration:
\begin{equation}
\biggl\langle \frac{n_i}{\Omega} \biggr\rangle - \biggl\langle \frac{n_i}{\Omega} \biggr\rangle_{F} = \Delta_i \Omega^{-2} + O(\Omega^{-5/2}).
\end{equation}
The relative error follows easily:
\begin{equation}
E_{mean}^i = \biggl[ \biggl\langle \frac{n_i}{\Omega} \biggr\rangle - \biggl\langle \frac{n_i}{\Omega} \biggr\rangle_{F} \biggr] \biggl \langle \frac{n_i}{\Omega} \biggr\rangle^{-1} = \frac{\Delta_i}{\phi_i} \Omega^{-2} + O(\Omega^{-5/2}).
\end{equation}
Similarly, using Eq. (22) and using the summary of results in Section III together with Eq. (52)  we find the absolute and relative errors in the variance of the fluctuations to respectively be given by:
\begin{align}
&\sigma_i^2 - \sigma_{F,i}^2 = \Delta_{ii} \Omega^{-2} + O(\Omega^{-5/2}), \\ 
&E_{var}^i = \frac{\sigma_i^2 - \sigma_{F,i}^2}{\sigma_i^2} = \frac{\Delta_{ii}}{\sigma_{i,LNA}^2} \Omega^{-2} + O(\Omega^{-5/2}),
\end{align}
where $\sigma_{i,LNA}^2$ is the variance in the concentration of species $i$ as estimated by the linear-noise approximation, i.e., $\sigma_{i,LNA}^2 = \Omega^{-1} ([\epsilon_i^2]_{0} - [\epsilon_i]_{0}^2)$. Hence the recipe for calculating the errors of the CFPE predictions is now complete. One first solves Eqs. (48-50) and then substitutes their solution in Eq. (54-57). Note that this calculation recipe is valid for all times and not only in steady-state conditions. 

Note also that since the denominator in Eq. (57) is the linear-noise approximation estimate for the variance then the leading relative error term in the variance is proportional to $\Omega^{-1}$. In contrast the leading relative error term in the mean concentrations, Eq. (55), is proportional to $\Omega^{-2}$. Hence the CFPE's estimates of mean concentrations are generally expected to be more accurate than those of the variance of the fluctuations about the mean concentrations.

Finally we obtain the absolute error in the CFPE prediction of the skewness of the probability distribution of the concentration of species $i$. The skewness is defined as:
\begin{equation}
s_i = \biggl \langle \biggl(\frac{n_i}{\Omega} - \biggl\langle \frac{n_i}{\Omega} \biggr\rangle \biggr)^3 \biggr \rangle \ \sigma_i^{-3}.
\end{equation}
The absolute error in the skewness is then $E_{skew}^i = s_i - s_{F,i}$ where $s_{F,i}$ is the skewness predicted by the CFPE, i.e., Eq. (58) with subscript $F$ throughout. As before, by using using Eqs. (21-22) together with the summary of results in Section III and Eq. (53) we get:
\begin{equation}
E_{skew}^i = \frac{\Delta_{iii}}{\sigma_{i,LNA}^3} \Omega^{-2} + O(\Omega^{-5/2}).
\end{equation}

\subsection{The CFPE is more accurate than the linear noise approximation}

We can now answer the question: which of the two, CFPE or linear Fokker-Planck equation, is the most accurate? We note that the linear Fokker-Planck equation (or equivalently the linear-noise approximation) is obtained by keeping only terms of order $\Omega^0$ in Eq. (14). If we do the same on the expansion of the CFPE, i.e. Eq. (36), then we also get the same linear Fokker-Planck equation. This equality implies that the CFPE becomes correct for large enough volumes or equivalently for large enough molecular populations. This result was known to van Kampen and is discussed in the book by Gardiner \cite{Gardiner}.

Within the linear-noise approximation, one can calculate the two quantities $[\epsilon_r ]_{0}$ and $[\epsilon_r \epsilon_k ]_{0}$ using Eqs. (26) and (23) respectively. The quantities $[\epsilon_r ]_{m}$ and $[\epsilon_r \epsilon_k ]_{m}$ where $m > 0$ are all zero in this approximation since the expansion has only terms to order $\Omega^0$. Now the initial condition for the CME is a delta function centered on the number of molecules as given by the REs, i.e. at time $t = 0$, the average number of molecules of the CME and the REs agree and hence it follows that $[\epsilon_r ]_{0} = 0$ initially and for all times \cite{vanKampen}. These results together with Eqs. (21) and (22), would seem to imply that within the linear-noise approximation, the mean concentrations are accurate to order $\Omega^{-1/2}$ while the variance is accurate to order $\Omega^{-1}$. However by considering terms of higher order than those leading to the linear-noise approximation, one arrives at the conclusion that actually the variance within this approximation is accurate to higher order than $\Omega^{-1}$.
This can be deduced by noting that $[\epsilon_r ]_{0} = 0$ for all times implies $[\epsilon_w \epsilon_k \epsilon_l]_{0} =[\epsilon_r \epsilon_k ]_{1}= 0$ also for all times. Hence it follows from Eq. (22) that the variance in the linear-noise approximation is accurate to order $\Omega^{-3/2}$.

Now from Eqs. (54) and (56), it is evident that generally the mean concentration and variance prediction of the CFPE are accurate to at least order $\Omega^{-3/2}$. Hence the mean concentration prediction of the CFPE is more accurate than that which can be obtained from the linear Fokker-Planck equation. It is also clear that the higher accuracy comes from taking into account the non-linear character of the CFPE since the $\Delta_i$ term in Eq. (54) is obtained by considering terms in Eqs. (14) and (36) of higher order than the linear-noise approximation. 

We can also derive an explicit equation for the mean concentrations predicted by the CFPE accurate to order $\Omega^{-1}$:
\begin{align}
\partial_t \biggl \langle \frac{n_i}{\Omega} \biggr \rangle_F &= \partial_t \phi_i + \Omega^{-1/2} (\partial_t [\epsilon_i]_{F,0} \Omega^0 + \partial_t [\epsilon_i]_{F,1} \Omega^{-1/2}) + O(\Omega^{-3/2}) \nonumber \\
& = \partial_t \phi_i + J_i^w \biggl(\biggl \langle \frac{n_i}{\Omega} \biggr \rangle_F - \phi_i \biggr) + \frac{1}{2} \Omega^{-1} (J_i^{wp} [\epsilon_w \epsilon_p]_{F,0} - J_i^{w(2)} \phi_w) + O(\Omega^{-3/2}).
\end{align}  
Note that the first step proceeds by taking the time derivative of Eq. (21) and the second step follows from using Eqs. (26) and (43), bearing in mind that $[\epsilon_i]_{F,0} = [\epsilon_i]_{0}$. Hence the computation of the mean concentrations to this order requires only the solution of the REs and of the Lyapunov equation Eq. (23). Note that Eq. (60) is exactly the same as the Effective Rate Equations recently derived by Grima from the CME (Eq. 60 is the same as Eq. (22) together with Eq. (24) in Ref \cite{Grima2010}). 

\subsection{The CFPE is highly accurate for equal-step reactions involving one species}

Consider the case where we have $N$ species interacting via $R$ elementary reactions of the equal-step type, i.e., in each individual reaction, either $p$ molecules of a species are generated or $p$ molecules are destroyed or no molecules are generated or destroyed. In such a case, the stoichiometric matrix elements are $S_{ij}=\pm p$ or $0$, where $p$ is a non-zero positive integer. Three examples of equal-step reactions are:
\begin{align}
&\O \xrightarrow{k_1} X_1, A + X_1 \xrightleftharpoons[k_3]{k_2}  2X_1, X_1 \xrightarrow{k4} \O \nonumber \\
&\O \xrightarrow{k_1} 2 X_1 \xrightarrow{k_2} \O \nonumber \\
&\O \xrightleftharpoons[k_2]{k_1} X_1, \O \xrightleftharpoons[k_4]{k_3} X_2, X_1+X_2 \xrightarrow{k_5} \O 
\end{align}
The first reaction is autocatalytic where $A$ is some very abundant species whose number of molecules is considered constant; this is a one-step, one species reaction scheme. The second reaction involves the burst input of two molecules and their dimerization, a two-step one species reaction scheme. The third reaction involves the production and degradation of two species and their bimolecular interaction; this is a one-step, two species reaction scheme.

For equal-step reactions, one species reaction schemes, the quantity $D_{111}$ evaluates to zero in steady-state conditions:
\begin{align}
D_{111} &= \sum_{j=1}^R (S_{1j})^3 f_j(\phi_1) \nonumber \\
 &= p^2 \sum_{j=1}^R S_{1j} f_j(\phi_1) = 0.
\end{align} 
Note that in the last step, use was made of the steady-state condition: $\partial_t \phi_1 = \sum_{j=1}^R S_{1j} f_j(\phi_1) = 0$. From Eqs. (48-50), we can then deduce that $\Delta_1 = \Delta_{11}=\Delta_{111}=0$. Hence it follows from Eqs. (54) and (56) that the mean concentrations and the variance of fluctuations predicted by the CFPE for one species, equal-step reactions, are accurate to at least order $\Omega^{-2}$. This is impressive when one considers that the linear-noise approximation of the CME only leads to estimates accurate to order $\Omega^{-1/2}$ in the mean and order $\Omega^{-3/2}$ in the variance. These conclusions lend support to the results of an early investigation of the one species CFPE \cite{Gillespie1980}. 

However, this high accuracy of the CFPE is not generally true for multispecies equal-step reactions. For example, for the third reaction scheme in the examples considered above, one finds $D_{111}=D_{222}=0$ and $D_{112}=D_{121}=D_{211}=D_{221}=D_{212}=D_{122} = -k_5 \phi_1 \phi_2 \ne 0$. The non-zero values of $D_{ijk}$ for some index values implies that the mean and variance predictions of the CFPE in this case are accurate to order $\Omega^{-3/2}$.

\subsection{CFPE is highly accurate for multispecies reactions obeying detailed balance}

In the previous subsection we have seen how $D_{hkl}$ is zero for one species, one-step reaction schemes and how this leads to a particularly high accuracy in the predictions of the CFPE. We now want to find the condition which forces $D_{hkl} = 0$ for chemical reactions involving any number of species. Consider the case where all reactions are reversible. Since each reaction can be paired with its reverse, it follows that the formula for $D_{hkl}$ can then be written as:
\begin{align}
D_{hkl} &= \sum_{j=1}^R S_{hj} S_{kj} S_{lj} f_j(\vec{\phi}) \nonumber \\
&= \sum_{z=1}^{R/2} S_{hz+} S_{kz+} S_{lz+} f_{z+}(\vec{\phi}) + S_{hz-} S_{kz-} S_{lz-} f_{z-} (\vec{\phi}) \nonumber \\ &= \sum_{z=1}^{R/2} S_{hz+} S_{kz+}S_{lz+} [ f_{z+}(\vec{\phi}) - f_{z-} (\vec{\phi})],
\end{align}
where the subscripts $+$ and $-$ indicates quantities evaluated for the forward and backward reactions respectively. The reversibility condition imposes $S_{hz+}=-S_{hz-}$ and was used in deriving the last step. Furthermore, a system of reversible reactions will always reach chemical equilibrium and in such conditions the system is characterized by detailed balance, i.e., $f_{z+}(\vec{\phi}) = f_{z-}(\vec{\phi})$, the forward and reverse rates of each elementary reversible reaction balance \cite{Keizer}. Hence by Eq. (63), $D_{hkl} = 0$, in detailed balance conditions, and consequently by Eqs. (48-50) and Eqs. (54) and (56), the CFPE's predictions of mean and variance are accurate to order $\Omega^{-2}$. Equilibrium conditions always imply detailed balance and hence our results suggest that the size of the differences between the predictions of the CFPE and the CME increase with how far is the system from equilibrium. 

\section{Applications}

\subsection{Dimerization}

As a first application of our theory, we will estimate the relative errors in the CFPE predictions for a dimerization reaction. This is the simplest case of a bimolecular reaction mechanism. The main purpose of considering such a reaction is that both its CME and CFPE are exactly solvable and hence it provides us with a direct test of our expressions for the leading order error in the means and the variances as predicted by the CFPE.  The set of reactions under study are:
\begin{align}
&\O \xrightarrow{k_{1}} X, \nonumber \\
&X + X \xrightarrow{k_2} Y.
\end{align}
Monomers, denoted as $X$, are pumped into some compartment at a rate $k_1$. Pairs of monomers react with rate constant $k_2$ to form a dimer molecule, $Y$. The concentration of dimers increases with time however the concentration of monomers becomes constant after a short time, i.e. the monomers reach a steady-state. Since $Y$ is not involved in the reaction, the mathematical description is solely in terms of the number of molecules of the monomers for the CME and CFPE and in terms of the monomer concentration for the RE.

The CME, Eq. (2), for the dimerization reaction reads:
\begin{equation}
\partial_t P(n_1,t) = k_1 \Omega (E^{-1}_1-1) P(n_1,t) + \frac{k_2}{\Omega} (E^2_1 - 1) n_1 (n_1 - 1) P(n_1,t).
\end{equation}
 Multiplying the equation on both sides by $s^{n_1}$ and summing over $n_1$ from $0$ to infinity, we get the equivalent generating function equation:
\begin{equation}
\partial_t F(s,t) = k_1 \Omega (s - 1)F(s,t)+\frac{k_2}{\Omega} ( 1 - s^2) \frac{\partial^2 F(s,t)}{\partial s^2},
\end{equation}
where $F(s,t) = \sum_{n_1} s^{n_1} P(n_1,t)$. This partial differential equation is solved in the steady-state with boundary conditions $F(1)=1$ and $F(-1)=0$ \cite{Mazo1975} leading to:
\begin{equation}
F(s) = z^{1/2} \frac{I_1(4 X z^{1/2})}{I_1(4 X)},
\end{equation}
where $z = (1+s)/2$, $X = \Omega (k_1/ 2 k_2)^{1/2}$ and $I_n$ is the modified Bessel function of the first kind of order $n$. The mean concentration and variance of the concentration fluctuations about this mean according to the CME are then given by the following expressions:
\begin{align}
\bigg \langle  \frac{n_1}{\Omega} \bigg \rangle &= \Omega^{-1} \frac{\partial F(s)}{\partial s} \bigg|_{s=1} = \frac{\phi_1 I_0(4 n_{ode}) }{ I_1 (4 n_{ode})}, \\
\sigma_1^2 &= \Omega^{-2} \biggl( \frac{\partial^2 F(s)}{\partial s^2} \bigg|_{s=1} + \frac{\partial F(s)}{\partial s} \bigg|_{s=1} - \biggl[\frac{\partial F(s)}{\partial s} \bigg|_{s=1} \biggr]^2 \biggr) \nonumber \\ &= \frac{\phi_1^2 [ n_{ode} [I_1(4 n_{ode})]^2 - n_{ode} [I_0(4 n_{ode})]^2 + I_0(4 n_{ode}) I_1(4 n_{ode}) ] }{n_{ode} [I_1(4 n_{ode})]^2}.
\end{align}
Note that these expressions are obtained within an exact approach and are not approximations as the ones stemming from the system-size expansion of the CME.  

Now we obtain expressions for the mean and variance using the CFPE approach. The CFPE, Eq. (33), for the dimerization reaction reads:
\begin{equation}
\frac{\partial P(n_1,t)}{\partial t} = \biggl( -\frac{\partial}{\partial n_1} A(n_1) + \frac{1}{2} \frac{\partial^2}{\partial n_1^2} B(n_1) \biggr) P(n_1,t),
\end{equation}
where $A = k_1 \Omega - 2 k_2 n_1 (n_1 - 1)/\Omega$ and $B = k_1 \Omega + 4 k_2 n_1 (n_1 - 1)/\Omega$.  The exact stationary solution of this non-linear second order partial differential equation can be shown to be:
\begin{align}
P(n_1) & = \frac{\exp \biggl[  -n_1 + \frac{3 k_1 \Omega^2 \arctan H(n_1)}{2 \sqrt{k_2} \sqrt{k_1 \Omega^2 - k_2}} \biggr]}{4 k_2 (n_1 - 1) n_1 + k_1 \Omega^2} \biggl (K_1 + K_2 \int_1^{n_1} d \eta \exp \biggl[ - \frac{3 k_1 \Omega^2 \arctan H(\eta)}{2 \sqrt{k_2} \sqrt{k_1 \Omega^2 - k_2}} + \eta \biggr] \biggr),
\end{align}
where $H(x) = \sqrt{k_2} (2 x - 1)/ \sqrt{k_1 \Omega^2 - k_2}$. The constants $K_1$ and $K_2$ are to be determined by the boundary conditions and the normalization condition. The boundary conditions of the CFPE are $P(n_1 = \pm \infty) = 0$. Note that the CFPE unlike the CME does not generally have a natural boundary at $n_1 = 0$ since the noise can sometimes drive the system to negative values of $n_1$ \cite{Hanggi1983}. Note that this problem is also implicit in the stationary solution of the linear Fokker-Planck equation, a Gaussian which is non-zero for negative particle numbers \cite{vanKampen} (see the end of this subsection for a further discussion of boundary conditions). The condition at $-\infty$ fixes the value of $K_2$ while the condition at $\infty$ is automatically satisfied by the exponential pre-factor. The remaining constant $K_1$ is fixed by the normalization condition. Since there is no closed form solution for the integral in Eq. (71), $K_1$ has to be computed numerically; once $P(n_1)$ is determined, the mean and variance can be straightforwardly numerically computed as well.  

The exact relative error in the mean and variance predictions of the CFPE can now be computed. One first fixes the rate constants $k_1$ and $k_2$ and $n_{ode}$. The normalization constant $K_1$ is found by numerical integration and from the ensuing steady-state probability distribution, one finds the mean, $\langle  n_1 \rangle / \Omega_{CFPE}$ , and variance $\sigma_{1,CFPE}^2$. The numerical error in the integration is essentially eliminated by performing the integration for a set of decreasing step size values and extrapolating to obtain the integral value at zero step size. Using the same values of rate constants and $n_{ode}$, one uses Eqs. (68-69) to compute the mean and variance according to the CME. The exact relative errors in the mean and variance can then be found using $1 - (\langle  n_1 \rangle / \Omega_{CFPE}) / (\langle  n_1 \rangle / \Omega)$ and $1 - \sigma_{1,CFPE}^2 / \sigma_{1}^2$, respectively. The exact absolute values of the relative errors in the CFPE predictions are shown by the red open circles in Fig. 1 for parameter values $k_1 = 1$ and $k_2 = 2$. Note that the relative error in the variance is larger than that in the mean. The errors increase with decreasing steady-state numbers of monomers. Even for very small numbers, the errors are quite small. For example for a case in which the REs predict 5 monomers in steady-state, the percentage relative errors in the mean and variance predictions of the CFPE are just $0.5 \%$ and $6.5\%$ respectively. The high accuracy of the CFPE in low particle number conditions is indeed surprising since typically it has only been deemed accurate for systems characterized by large particle numbers. 

We can now test the accuracy of the theory developed in the previous sections by using it to obtain expressions for the approximate relative errors in the mean and variance and then compare these with the exact values as already obtained above. By inspection of the reaction scheme, Eq. (64), it can be easily deduced that the stoichiometric matrix is $S = (1, -2)$. From the definition of the macroscopic rate function vector (see Introduction) it also follows that it is equal to $\vec{f}(\phi_1) = (k_1, k_2 \phi_1^2)$. This is all the information needed to calculate the estimates for the relative errors using our theory. The macroscopic concentration and the relevant entries of the D and J matrices evaluated at steady-state are then given by:
\begin{align}
&\phi_1 = \biggl(\frac{k_1}{2 k_2} \biggr)^{1/2}, \\
&D_{11} = \sum_{j=1}^2 (S_{1j})^2 f_j(\phi_1) = k_1 + 4 k_2 \phi_1^2, \quad D_{111} = \sum_{j=1}^2 (S_{1j})^3 f_j(\phi_1) = k_1 - 8 k_2 \phi_1^2, \\
&J_1^1 = \frac{\partial}{\partial \phi_1}  \sum_{j=1}^2 S_{1j} f_j(\phi_1) = -4 k_2 \phi_1, J_1^{11}=\frac{\partial}{\partial \phi_1} J_1^1 = -4 k_2.
\end{align}
These are substituted in Eqs. (48-50) which are then evaluated at steady-state, leading to:
\begin{align}
&\Delta_{111} = -\frac{D_{111}}{3 J_1^1} = - \frac{1}{2} \phi_1, \\
&\Delta_{11} = -\frac{J_{1}^{11} \Delta_{111}}{2 J_1^1} = \frac{1}{4}, \\ 
&\Delta_{1} = -\frac{J_{1}^{11} \Delta_{11}}{2 J_1^1} = -\frac{1}{8 \phi_1}.
\end{align}
The relative error in the mean concentration to leading order is then given by Eq. (55):
\begin{equation}
E_{mean}^1 = -\frac{1}{8 n_{ode}^2},
\end{equation}
where $n_{ode} = \Omega \phi_1$ is the average number of monomers as predicted by the REs. To compute the relative error in the variance we need to first estimate the variance to the linear-noise level of approximation. This is done by solving Eq. (23) in steady-state:
\begin{equation}
[\epsilon_1^2]_0 = -\frac{D_{11}}{2 J_1^1} = \frac{3 k_1}{8 k_2 \phi}.
\end{equation} 
The variance is then $\sigma_{1,LNA}^2 = \Omega^{-1} [\epsilon_1^2]_0$. Using the latter and Eq. (76), it is found that Eq. (57) evaluates to:
\begin{equation}
E_{var}^1 = \frac{1}{3 n_{ode}}.
\end{equation}

The theoretical absolute values of the relative errors in the CFPE predictions, as given by Eq. (78) and Eq. (80), are shown by the solid blue lines in Fig. 1 for parameter values $k_1 = 1$ and $k_2 = 2$. The theory is generally in very good agreement with the exact solution; small discrepancies are only apparent in the error for the variance at molecule numbers less than approximately 5 monomers. The comparison has also been done for many other parameters values and as predicted by theory, in all cases, the graphs are the same as shown in Fig. 1. 

We have also computed the exact errors by solving the CFPE with different boundary conditions. One could argue that constraints should be imposed on the CFPE such that it preserves the natural boundary of the CME at $n_1 = 0$. This can be fulfilled by requiring that the probability current of the CFPE vanishes at $n_1 = 0$ \cite{Risken}. In such a case the stationary solution of the CFPE has the form of Eq. (71) with $K_2 = 0$ and $K_1$ is found by requiring that the solution is normalized on $(0,\infty)$. The exact errors computed with this new solution of the CFPE are practically indistinguishable from the previous solutions shown in Fig. 1 except for a small discrepancy at $n_{ODE} = 3$. The excellent agreement of our theoretical solution with both CFPE solutions is simply due to the fact that the probability of $n_1$ taking negative values in Eq. (71) is very small, unless $n_{ode}$ is also very small. 

\subsection{Enzyme catalysis: the Michaelis-Menten mechanism}
As a second application, we consider the catalysis of a substrate species $S$ into a product species $P$ by an enzyme species via the Michaelis-Menten mechanism \cite{CornishBowden1995}:
\begin{align}
&\O \xrightarrow{k_{in}} S, \quad
S + E \xrightleftharpoons[k_1]{k_0}C, \\ &C \xrightarrow{k_2} E + P,
\end{align}
where $E$ denotes the free enzyme, i.e. when it is not bound to substrate, and $C$ denotes the substrate-enzyme complex. We will denote substrate, complex and free enzyme as species 1, 2 and 3 respectively. Note that the product species is missing from the kinetic description because it is a byproduct of the reaction and thus not involved in the reactions. The total enzyme concentration is a constant, $\phi_2 + \phi_3 = \langle n_2/\Omega \rangle + \langle n_3/\Omega \rangle = E_T$, since the enzyme is either bound to substrate or unbound. Hence we effectively have a two variable system. The reaction system exhibits a steady-state in the concentrations of substrate and complex whenever the inequality $k_{in} \le k_2 E_T$ is satisfied, i.e. when the rate at which substrate is pumped into the system is less than or equal to the maximum rate at which the enzyme can convert substrate to product. Assuming such conditions, our aim is to calculate the relative errors in the mean and variance predictions of the CFPE, i.e. Eqs (55) and (57); to achieve this, we will first need to solve Eqs. (48)-(50), which we show in detail now.    

The stoichiometric matrix and the macroscopic rate function vector follow directly from their definitions (see Introduction):
\begin{align}
 \textbf{S} =  \begin{pmatrix} 1 & -1 & 1 & 0 \\ 0 & 1 & -1 & -1\end{pmatrix}, \quad
 \vec{f}(\phi_1,\phi_2) = \{ k_{in}, k_0 (E_T - \phi_2) \phi_1, k_1 \phi_2, k_2 \phi_2\} \nonumber.
\end{align}
The rate equations and the $D$ and $J$ matrices follow by inserting the above in Eq. (1), Eq. (16) and Eq. (17) to obtain:
\begin{align}
&\phi_1= K_M \frac{1 - \beta}{\beta}, \quad \phi_2 = E_T (1 - \beta), \\
&J_1^1  =-k_0 (E_T - \phi_2), J_1^2 = k_1 + k_0 \phi_1, J_2^1 = -J_1^1, J_2^2=-(k_1+k_2+k_0\phi_1) \\
&J_{1}^{11} = J_{1}^{22}=J_2^{11}=J_{2}^{22}=0, J_{1}^{12} = J_{1}^{21}=-J_2^{12}=-J_{2}^{21}=k_0, \\
&D_{111}=D_{222}=0, \\ &D_{112}=D_{121}=D_{211}=-D_{122}=-D_{212}=-D_{221}= E_T (k_1+k_2) \eta (1 - \beta), \\
&D_{11}=D_{22}=2 E_T (k_1+k_2) (1 - \beta), D_{12}=D_{21}=E_T (k_1+k_2) (1 - \beta)(\eta - 2),
\end{align}
where $\beta = 1 - k_{in}/(k_2 E_T)$, $K_M = (k_1+k_2)/k_0$ is the Michaelis-Menten constant and $\eta = 1 - k_1/(k_0 K_M)$. 

Note that $\beta$ is a measure of enzyme saturation since as the input rate of substrate, $k_{in}$, approaches the maximum rate at which the enzyme can catalyze the reaction, $k_2 E_T$, the proportion of enzyme in complex form increases accordingly, as can also be seen from Eq. (83). Note also that $\eta$ is a measure of how far is the system from equilibrium. This is since if substrate binding would occur at equilibrium, i.e., $k_{in}=k_2=0$, then the relationship between the macroscopic concentrations would be $\phi_1 \phi_3 / \phi_2 = k_1/k_0$ while generally in steady-state conditions, i.e.  $k_{in}>0,k_2>0,\beta \le 1$, the relationship between the macroscopic concentrations is $\phi_1 \phi_3 / \phi_2 = K_M$.  Both $\beta$ and $\eta$ are non-dimensional, positive fractions. 
 
Substituting Eqs. (84-87) in Eqs. (48-50), setting the time derivative to zero and solving the resulting set of simultaneous equations we obtain:
\begin{align}
&\Delta_1 = \frac{-(1 - \beta) \beta \eta}{
 2 K_M (1 + u \beta^2)^3 + E_T \beta^3 (1 + u \beta^2) \eta},
 \quad \Delta_2 = 0, \\
&\Delta_{11} =  \frac{(1 - \beta) \eta (1 + \beta (u \beta (3 + 
          u \beta^2) + \eta))}{
 2 (1 + u \beta^2)^3 + u \beta^3 (1 + u \beta^2) \eta}, \\
&\Delta_{12} = \Delta_{21} = -\Delta_{22} = -\frac{u (1 - \beta) \beta^2 \eta}{
2 (1 + u \beta^2)^3 + u \beta^3 (1 + u \beta^2) \eta},
\end{align}
where $u = E_T / K_M$.  

The leading order term of the relative errors in the mean concentrations of substrate and complex, as predicted by the CFPE, are then given by substituting Eq. (83) together with Eq. (89) in Eq. (55):
\begin{equation}
E_{mean}^1=\frac{-\beta^2 \eta}{K_M^2 \Omega^2 (1 + u \beta^2) (2 + u \beta^2 (4 + \beta (2 u \beta + \eta)))}, \quad E_{mean}^2=0.
\end{equation}
Note that the relative error in the substrate concentration is always negative, i.e., the CFPE overestimates the mean substrate concentrations and it increases with the distance from equilibrium, $\eta$. There is no error in the CFPE estimate for enzyme concentration (at least to order $\Omega^{-2}$).

To calculate the relative errors in the variance using Eq. (57) we first need to compute the variance as estimated by the linear-noise approximation. This is obtained by solving Eq. (23) using Eq. (84) and Eq. (88):
\begin{equation}
\sigma_{1,LNA}^2=\frac{K_M (1 - \beta) (1 + 
    u \beta^3 + (\beta - 1) \beta \eta)}{\beta^2 \Omega (1 + 
    u \beta^2)}, \quad \sigma_{2,LNA}^2 = \frac{E_T (1 - \beta) \beta (1 + u \beta)}{\Omega(
 1 + u \beta^2)}.
\end{equation}
Finally substituting the above two equations and Eqs. (90-91) in Eq. (57) we obtain the leading order term of the relative errors in the variance of the substrate and complex concentration fluctuations, as predicted by the CFPE:
\begin{align}
E_{var}^1&=\frac{\beta^2 \eta (1 + \beta (u \beta (3 + 
         u \beta^2) + \eta))}{K_M \Omega (1 + 
   u \beta^3 + (\beta - 1) \beta \eta) (2 + 
   u \beta^2 (4 + \beta (2 u \beta + \eta)))}, \\
E_{var}^2&= \frac{\beta \eta}{K_M \Omega (1 + u \beta) (2 + 
   u \beta^2 (4 + \beta (2 u \beta + \eta)))}.
\end{align}
Note that both relative errors are always positive implying that the CFPE underestimates the variance.

We can now use the formulae given by Eq. (92), Eq. (94) and Eq. (95) to estimate the relative error of the CFPE when modeling conditions typical of the intracellular environment. A principal characteristic of such an environment is that the number of molecules of some species can be quite small. A detailed protein abundance profiling of the Escherichia coli cytosol by Ishihama et al \cite{Ishihama2008} shows that the total number of enzyme molecules per cell approximately varies from a hundred to a few thousands. It is indeed in this limit of small numbers that it is frequently thought that the CFPE and the CLE description are not very accurate. We quantitatively test this hypothesis using our formulae. We will first express our error formulae in terms of the average number of molecules of substrate and free enzyme as predicted by the REs, i.e., $n_{1,ODE} = \phi_1 \Omega$ and $n_{3,ODE} = \phi_3 \Omega$. Using Eqs (83) we find that:
\begin{align}
K_M \Omega = \frac{\beta n_{1,ODE}}{1 - \beta}, \quad
u = \frac{1 - \beta}{\beta^2} \frac{n_{3,ODE}}{n_{1,ODE}}.
\end{align}
Substituting Eq. (96) in Eq. (92), Eq. (94) and Eq. (95) we get expressions for the errors in terms of $n_{1,ODE}$, $n_{3,ODE}$, $\beta$ and $\eta$. Given fixed molecule numbers, $n_{1,ODE}$ and $n_{3,ODE}$, we can find the maximum error by varying $\beta$ and $\eta$ over their allowed range $[0,1]$. Repeating this procedure for various molecule numbers we can obtain simple two dimensional plots of the maximum error. The results for the maximum relative error in the predictions of the variance are shown in Fig. 2.  The results verify that the predictions of the CFPE become increasingly accurate with increasing molecule numbers. They also show that the error incurred by using the CFPE for cases of small molecule numbers is very small: less than $1\%$ for a few tens of molecule numbers. 

It is noteworthy that this accuracy is far better than even that hypothesized by proponents of the CFPE \cite{Gillespie2000}. For example Gillespie in his seminal paper on the derivation of the CLE  \cite{Gillespie2000} remarks in his conclusion that the CLE (and hence the CFPE) approximation is probably not a good one when one models a system composed of three time-varying species with total molecular population of 2000 since it appears quite possible that the molecule number of at least one of the species becomes significantly small at some point in time. In contrast our theory seems to predict that the CFPE predictions will still be very accurate even when the molecule numbers are quite low. 

We have tested these predictions by numerically solving the CLE for the Michaelis-Menten process using the Euler-Mayurama method to obtain the mean substrate concentrations and the variance of the substrate fluctuations about the means. The same were obtained from stochastic simulation algorithm simulations of the CME. The results are shown in Fig. 3. The parameters are chosen to be $k_0=272$, $k_1=8$, $k_2=60$, $E_T=100$, $k_{in}=5880$ and $\Omega=25$ since this gives conditions similar to those mentioned by Gillespie above. The RE solutions, Eqs. (83), with the above parameters  lead to $\phi_1 = 12.25$, $\phi_2 = 98$ and $\phi_3 = 2$ which, given a volume of $\Omega=25$, would imply $n_{1,ODE} = 306.25$, $n_{2,ODE} = 2450$ and $n_{3,ODE} = 50$. The total molecular population of enzyme (free plus complex form) is 2500 molecules. Each algorithm (Euler-Mayurama and stochastic simulation algorithm) was run 5 times leading to 5 independent estimates \cite{Samplingproc}. Note that even though the mean number of free enzyme molecules is considerably low, the predictions of the CFPE for both the mean and the variance agree (within sampling error) with those of the CME. For comparison we have also plotted the predictions of the linear-noise approximation (red lines) and of the mean concentration as predicted by the Effective Mesoscopic Rate Equation Eq. (60) (blue line). The results clearly confirm that the CFPE is more accurate than the linear Fokker Planck equation associated with the linear-noise approximation and that indeed the mean concentrations of the CFPE are in excellent agreement with the Effective Mesoscopic Rate Equations derived in Ref \cite{Grima2010}. The Effective Mesoscopic Rate Equation for the Michaelis-Menten reaction was first obtained in Ref \cite{Grima2009} (See Eq. (29) in the latter reference). 

For our set of parameters, the theoretical expressions, Eqs. (92) and (94), evaluate to $E_{mean}^1 =  -2.9 \times 10^{-6}$ and $E_{var}^1 = 3.2 \times 10^{-5}$; these errors are so small that they are clearly  masked by the sampling error inherent in the calculation of the mean and the variance from the long-time simulation trajectories. Indeed, in agreement with our theory, from Fig. 3 one can detect no significant difference between the CFPE and CME predictions. The numerical experiments were performed with various other parameter sets -- in all cases we could not detect any discrepancy between the CFPE and CME predictions within sampling error.

\section{Discussion and Conclusion}

Summarizing, in this article we have shown that (i) the mean and variance predictions of the CFPE are accurate to order $\Omega^{-3/2}$. Since those of the linear Fokker-Planck equation are accurate to order $\Omega^{-1/2}$ for the mean and $\Omega^{-3/2}$ for the variance, it is clear that the CFPE is generally more accurate than the linear Fokker-Planck equation or equivalently the linear-noise approximation. (ii) for detailed balance conditions, the predictions of the CFPE are even more accurate, order $\Omega^{-2}$, i.e. in equilibrium or near equilibrium conditions the CFPE does an excellent job of approximating the CME. (iii) accuracy to such high order in inverse powers of the system volume implies that the CFPE estimates should be quite good even for small populations of molecules. Our simulations for dimerization and enzyme-catalyzed reactions support these theoretical conclusions, with impressively good agreement down to an average of 5 molecules for the dimerization example.   

The CFPE's accuracy is indeed surprising given that it arises out of a naive truncation of the Kramers-Moyal expansion of the CME and that the CFPE cannot be obtained from the systematic system-size expansion of the CME. Only the linear Fokker-Planck equation can be derived from the latter expansion by considering terms of order $\Omega^0$. This equation leads to mean and variance estimates which are accurate to orders $\Omega^{-1/2}$ and $\Omega^{-3/2}$. Now if one wants more accurate estimates one needs to consider higher-order terms in the expansion. To get mean concentration estimates to order $\Omega^{-1}$ one needs to consider the term in the system-size expansion proportional to $\Omega^{-1/2}$ \cite{Grima2010}. To this order, one does not obtain the CFPE, rather one obtains a partial differential equation with a third-order derivative. However, it turns out that the mean calculated from this equation precisely agrees with that calculated from the CFPE to order $\Omega^{-1}$.    
If we even wanted to get more accurate means and variance, say both to order $\Omega^{-2}$, we need to consider terms in the system-size expansion to order $\Omega^{-3/2}$. This leads to a partial differential equation for the time evolution of the probability density function with derivatives as high as fifth order. Once again this is not the CFPE. However under steady-state conditions obeying detailed balance, the estimates from this high-order differential equation and the CFPE exactly agree to order $\Omega^{-2}$. Hence we have shown that though it is true that the CFPE does not arise out of the system-size expansion, nevertheless its predictions are better than those which can be obtained by considering only the first term of the expansion (the linear-noise approximation) as is conventional \cite{ElfEhrenberg}. It follows that the non-linear character of the CFPE is not completely spurious as originally suggested by van Kampen \cite{vanKampen1982}. 

Our study is the first one to our knowledge which systematically analyzes the validity of the non-linear multivariate CFPE and which derives approximate expressions for the size of the errors in the CFPE estimates -- previous studies \cite{Gillespie2002,Higham2008} have focused on the CFPE for unimolecular reactions and for unimolecular and bimolecular reactions involving one species \cite{Gillespie1980}.  Our analysis is based on the system-size expansion and thus has the same limitations, namely that it is only applicable for chemical systems which are ``asymptotically stable in the sense of Lyapunov''. This implies that from our analysis we cannot draw any conclusions for bistable systems \cite{vanKampen}. Within these constraints, the system-size expansion is a legitimate means of obtaining the moments of the CME accurate to any desired order \cite{vanKampen}. A few authors \cite{Horsthemke} have expressed reservations regarding the accuracy of the expansion beyond the linear-noise level, their reasoning stemming from the fact that Pawula's theorem \cite{Pawula1967} states that a time-evolution equation for a probability density function with higher than second-order derivatives cannot describe a stochastic process. However these misgivings are undue -- the higher-order partial differential equation stemming from the expansion truncated to some order is ``not an exact equation for a Markov process that in some way approximates the original process; rather it is an approximate equation for the exact P.'' \cite{vanKampen1982}. This statement of van Kampen is generally true for any legitimate expansion of the CME, not only the system-size expansion; for example Risken and Vollmer \cite{Risken1979} showed that taking into account higher-order derivatives than two in the Kramers-Moyal expansion of the CME also leads to more accurate solutions than if one just had to use the CFPE. The accuracy of the system-size expansion beyond the linear-noise approximation has also been verified by many recent studies \cite{Grima2009PRL,Grima2009, Grima2010,Thomas2010,Fanelli1,Fanelli2}, putting at rest any small doubts about its general validity. Finally, the good agreement of our theoretical expressions for the errors with simulations is a clear indication of the soundness of our system-size expansion based approach.

Concluding our results offer theoretical and numerical support for Gillespie's hypothesis \cite{Gillespie2000} regarding the validity of the CFPE in both mesoscopic and macroscopic systems. Our formulas provide a simple means to estimate the error in the predictions of the CFPE and the associated CLE and hence should be of wide applicability to both theoretical and numerical studies of stochastic chemistry. 
 
\section*{Acknowledgments}
R. G. acknowledges support by SULSA (Scottish Universities Life Science Alliance).

\appendix

\section{Subtleties of the perturbative expansion in the probability density}

By the normalization condition and the expansion of $\Pi(\vec{\epsilon},t)$ we have:
\begin{equation}
\int \Pi(\vec{\epsilon},t) d\vec{\epsilon} = 1 = \sum_{j=0}^{\infty } \int \Pi_j(\vec{\epsilon},t) \Omega^{-j/2} d\vec{\epsilon}.
\end{equation}
Equating powers of the volume we obtain:
\begin{align}
\int \Pi_0 d\vec{\epsilon} &= 1, \\
\int \Pi_j d\vec{\epsilon} &= 0, \forall j \ge 1
\end{align}
An analogous property has been discussed by Gardiner in the different though related context of small noise expansions of the Fokker-Planck equation \cite{Gardiner}. The two properties above are useful in the computation of the integrals needed to arrive to Eqs. (23-32); for more details see Appendix B.

It follows from Eqs. (A2-A3) that only $\Pi_0$ is a genuine probability density while the higher orders are negative in some regions of the $\vec{\epsilon}$ space. From Eq. (15), we see that to order $\Omega^0$, the time-evolution of $\Pi_0$ is given by a linear Fokker-Planck equation:
\begin{equation}
\frac{\partial \Pi_0(\vec{\epsilon},t)}{\partial t} = -J_i^w \partial_i (\epsilon_w \Pi_0) + \frac{1}{2} D_{ip} \partial_{ip}^2 \Pi_0,
\end{equation}
which again verifies that $\Pi_0$ is a probability density. However the time-evolution equations for $\Pi_j$ where $j \ge 1$, involve derivatives of order larger than two and hence by Pawula's theorem \cite{Pawula1967} $\Pi_j$ cannot be genuine probability density functions. 

The above arguments also imply that it is not correct to think of $[ \epsilon_k \epsilon_m .. \epsilon_r ]_j$, where $j \ge 1$, as genuine statistical moments; rather they are best considered as placeholders or labels for the associated integrals $\int \epsilon_k \epsilon_m .. \epsilon_r \Pi_j d\vec{\epsilon}$. In the main text we refer to them as corrections to the moments to order $\Omega^{-j/2}$. It is however important to bear in mind that though $[ \epsilon_k \epsilon_m .. \epsilon_r ]_j$ are generally not true statistical moments, their linear superposition via Eq. (19) is a genuine statistical moment. Hence it is best to avoid associating any physical meaning to $[ \epsilon_k \epsilon_m .. \epsilon_r ]_j$ and to simply regard them as a means to obtain the desired answer, i.e., $\langle \epsilon_k \epsilon_m .. \epsilon_r \rangle$.

\section{Detailed derivation of the time-evolution equations for $[ \epsilon_r \epsilon_k ]_2$}

The time-evolution equations are obtained by substituting Eq. (18) in Eq. (15), multiplying the resulting equation on both sides by $ \epsilon_r \epsilon_k$ and integrating over $d\vec{\epsilon}$. Finally we equate terms of order $\Omega^{-1}$ on both sides of the equation to obtain the time-evolution equation for $[ \epsilon_r \epsilon_k ]_2$. The right hand side of the resulting equation simplifies by performing integration by parts; there are 8 integrals which need such evaluation and we treat each one of them below. 

\begin{enumerate}

\item 
\begin{align}
J_{i}^w \int \epsilon_r \epsilon_k \partial_i (\epsilon_w \Pi_2) d{\vec{\epsilon}} &= - J_{i}^w \int \epsilon_w \Pi_2 [\epsilon_k \delta_{i,r} + \epsilon_r \delta_{i,k}] d{\vec{\epsilon}} \nonumber \\ &= - J_{i}^w ([ \epsilon_w \epsilon_k ]_2 \delta_{i,r} + [ \epsilon_w \epsilon_r ]_2 \delta_{i,k}) \nonumber \\& =  - J_{r}^w [ \epsilon_w \epsilon_k ]_2 - J_{k}^w [ \epsilon_w \epsilon_r ]_2.
\end{align}
Note that in Eq. (15) we are summing over all twice repeated indices, which for the above integral are $i$ and $w$. Use was made of this implicit summation on $i$ in the derivation of the last step.

\item
\begin{align}
D_{ip} \int \epsilon_r \epsilon_k \partial_{ip} (\Pi_2) d{\vec{\epsilon}} &= - D_{ip} \int \partial_p \Pi_2 [\epsilon_k \delta_{i,r} + \epsilon_r \delta_{i,k}] d{\vec{\epsilon}} \nonumber \\ &= D_{ip}(\delta_{p,r} \delta_{i,k} + \delta_{p,k} \delta_{i,r} ) \int \Pi_2 d{\vec{\epsilon}} = 0. 
\end{align}
In the last step, we have made use of the fact that $\int \Pi_2 d{\vec{\epsilon}} = 0$, as shown in Appendix A.

\item
\begin{align}
J_{i}^{wp} \int \epsilon_r \epsilon_k \partial_i (\epsilon_w \epsilon_p \Pi_1) d{\vec{\epsilon}} &= - J_{i}^{wp} \int \epsilon_w \epsilon_p \Pi_1 [\epsilon_k \delta_{i,r} + \epsilon_r \delta_{i,k}] d{\vec{\epsilon}} \nonumber \\ &= - J_{i}^{wp} ([ \epsilon_w \epsilon_p \epsilon_r ]_1 \delta_{i,k} + [ \epsilon_w \epsilon_p \epsilon_k ]_1 \delta_{i,r}) \nonumber \\& =  - J_{k}^{wp} [ \epsilon_w \epsilon_p \epsilon_r ]_1 - J_{r}^{wp} [ \epsilon_w \epsilon_p \epsilon_k ]_1.
\end{align}

\item
\begin{align}
J_{i}^{w(2)} \int \epsilon_r \epsilon_k \partial_i \Pi_1 d{\vec{\epsilon}} &= - J_{i}^{w(2)} \int \Pi_1 [\epsilon_k \delta_{i,r} + \epsilon_r \delta_{i,k}] d{\vec{\epsilon}} \nonumber \\ &= - J_{i}^{w(2)} ([ \epsilon_r ]_1 \delta_{i,k} + [ \epsilon_k ]_1 \delta_{i,r}) \nonumber \\& =  - J_{k}^{w(2)} [ \epsilon_r ]_1 - J_{r}^{w(2)} [ \epsilon_k]_1.
\end{align}

\item
\begin{align}
J_{ip}^w \int \epsilon_r \epsilon_k \partial_{ip} (\epsilon_w \Pi_1) d{\vec{\epsilon}} &= - J_{ip}^w \int \partial_p (\epsilon_w \Pi_1) [\epsilon_k \delta_{i,r} + \epsilon_r \delta_{i,k}] d{\vec{\epsilon}} \nonumber \\ &= J_{ip}^w [ \epsilon_w ]_1 (\delta_{p,r} \delta_{i,k} + \delta_{p,k} \delta_{i,r} ) \nonumber \\ &= 2 J_{kr}^{w} [ \epsilon_w ]_1.
\end{align}
In obtaining the last step we have used the implicit summation over $i$ and $p$ and also the symmetrical property, $J_{kr}^w=J_{rk}^w$, which follows from the definitions given by Eqs. (16-17).

\item
\begin{align}
J_{i}^{w(2)} \int \epsilon_r \epsilon_k \partial_i (\epsilon_w \Pi_0) d{\vec{\epsilon}} &= - J_{i}^{w(2)} \int \epsilon_w \Pi_0 [\epsilon_k \delta_{i,r} + \epsilon_r \delta_{i,k}] d{\vec{\epsilon}} \nonumber \\ &= - J_{i}^{w(2)} ([ \epsilon_w \epsilon_k ]_0 \delta_{i,r} + [ \epsilon_w \epsilon_r ]_0 \delta_{i,k}) \nonumber \\& =  - J_{r}^{w(2)} [ \epsilon_w \epsilon_k ]_0 - J_{k}^{w(2)} [ \epsilon_w \epsilon_r ]_0.
\end{align}

\item
\begin{align}
J_{ip}^{wm} \int \epsilon_r \epsilon_k \partial_{ip} (\epsilon_w \epsilon_m \Pi_0) d{\vec{\epsilon}} &= - J_{ip}^{wm} \int \partial_p (\epsilon_w \epsilon_m \Pi_0) [\epsilon_k \delta_{i,r} + \epsilon_r \delta_{i,k}] d{\vec{\epsilon}} \nonumber \\ &= J_{ip}^{wm} [ \epsilon_w \epsilon_m ]_0 (\delta_{p,r} \delta_{i,k} + \delta_{p,k} \delta_{i,r} ) \nonumber \\ &= 2 J_{kr}^{wm} [ \epsilon_w \epsilon_m ]_0.
\end{align}
Note that in the last step we have used the symmetrical property, $J_{kr}^{wm}=J_{rk}^{wm}$, which follows from the definitions given by Eqs. (16-17).

\item
\begin{align}
J_{ip}^{w(2)} \int \epsilon_r \epsilon_k \partial_{ip} (\Pi_0) d{\vec{\epsilon}} &= - J_{ip}^{w(2)} \int \partial_p \Pi_0 [\epsilon_k \delta_{i,r} + \epsilon_r \delta_{i,k}] d{\vec{\epsilon}} \nonumber \\ &= J_{ip}^{w(2)}(\delta_{p,r} \delta_{i,k} + \delta_{p,k} \delta_{i,r} ) \int \Pi_0 d{\vec{\epsilon}} = 2 J_{kr}^{w(2)}. 
\end{align}
In the last step, we have made use of the fact that $\int \Pi_0 d{\vec{\epsilon}} = 1$, as shown in Appendix A and the symmetry property used in the evaluation of the previous integral.

\end{enumerate}

\begin{figure}[ht]
\centering
\subfigure[]{ 
\includegraphics[width=3.0in]{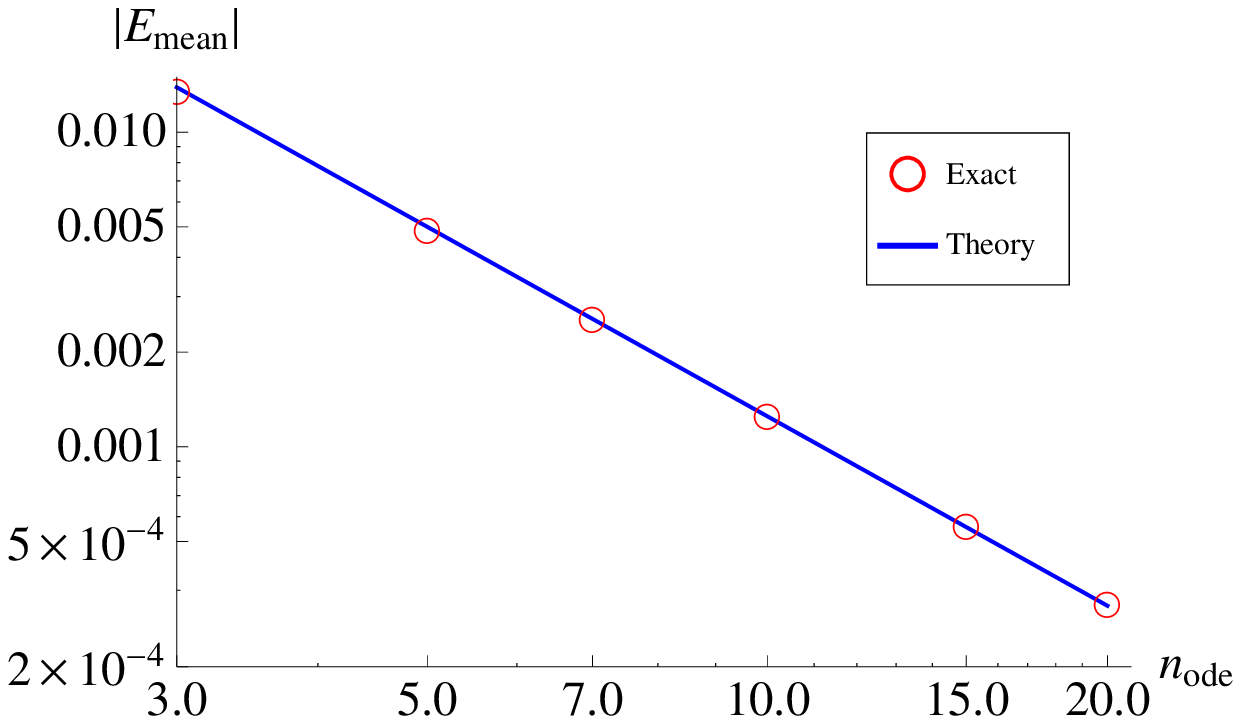}
\label{fig:subfig1}
}
\subfigure[]{
\includegraphics[width=3.0in]{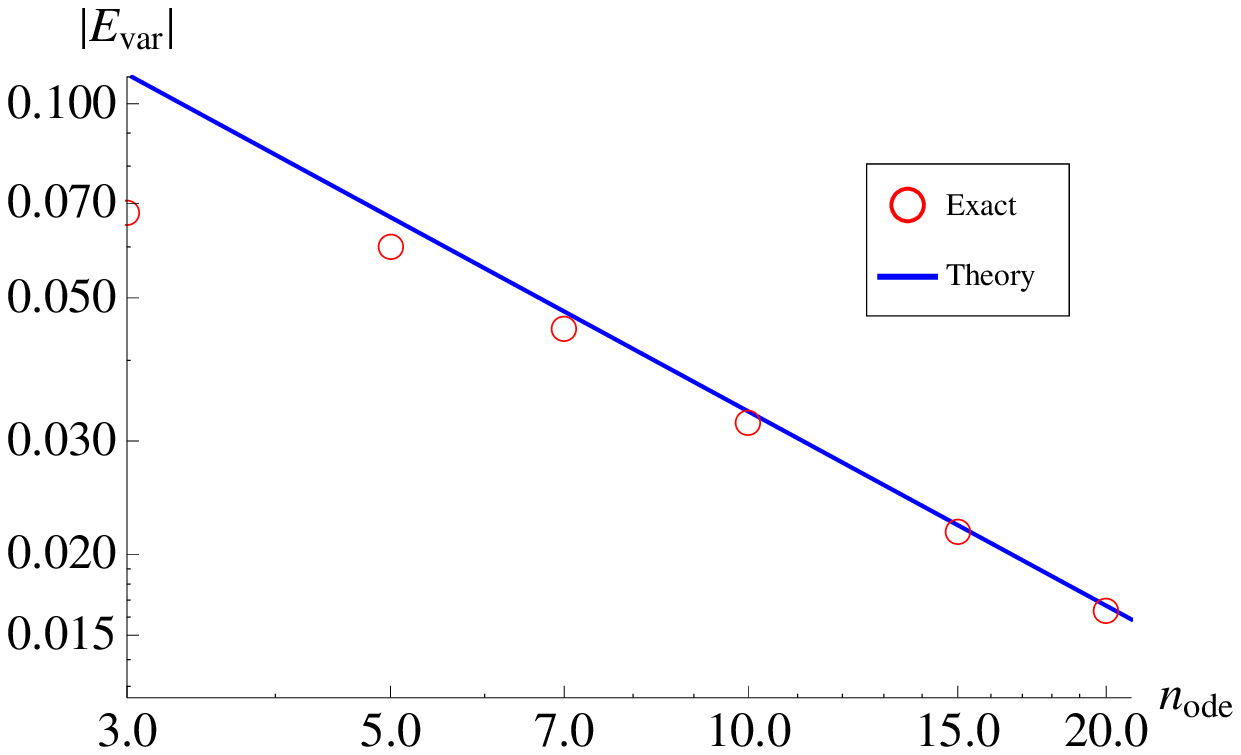}
\label{fig:subfig2}
}
\caption[]{Dependence of the absolute value of the relative errors in the CFPE prediction of the mean, $|E_{mean}|$ and variance, $|E_{var}|$, with the steady-state number of molecules, $n_{ode}$, as estimated by the rate equations. The red open circles show the errors computed using the exact solutions of the CFPE and the CME. The blue lines denote the leading order errors estimated by our theory and given by Eq. (78) in (a) and Eq. (80) in (b). Note that the leading order error estimates are in good agreement with the errors calculated from the exact solutions. Note also that the error made by the CFPE increases with decreasing molecule numbers and that the error in the variance is considerably larger than that in the mean, in many cases by more than one order of magnitude. See text for details and discussion.}
\end{figure}

\begin{figure}[ht]
\centering
\subfigure[]{ 
\includegraphics[width=3.0in]{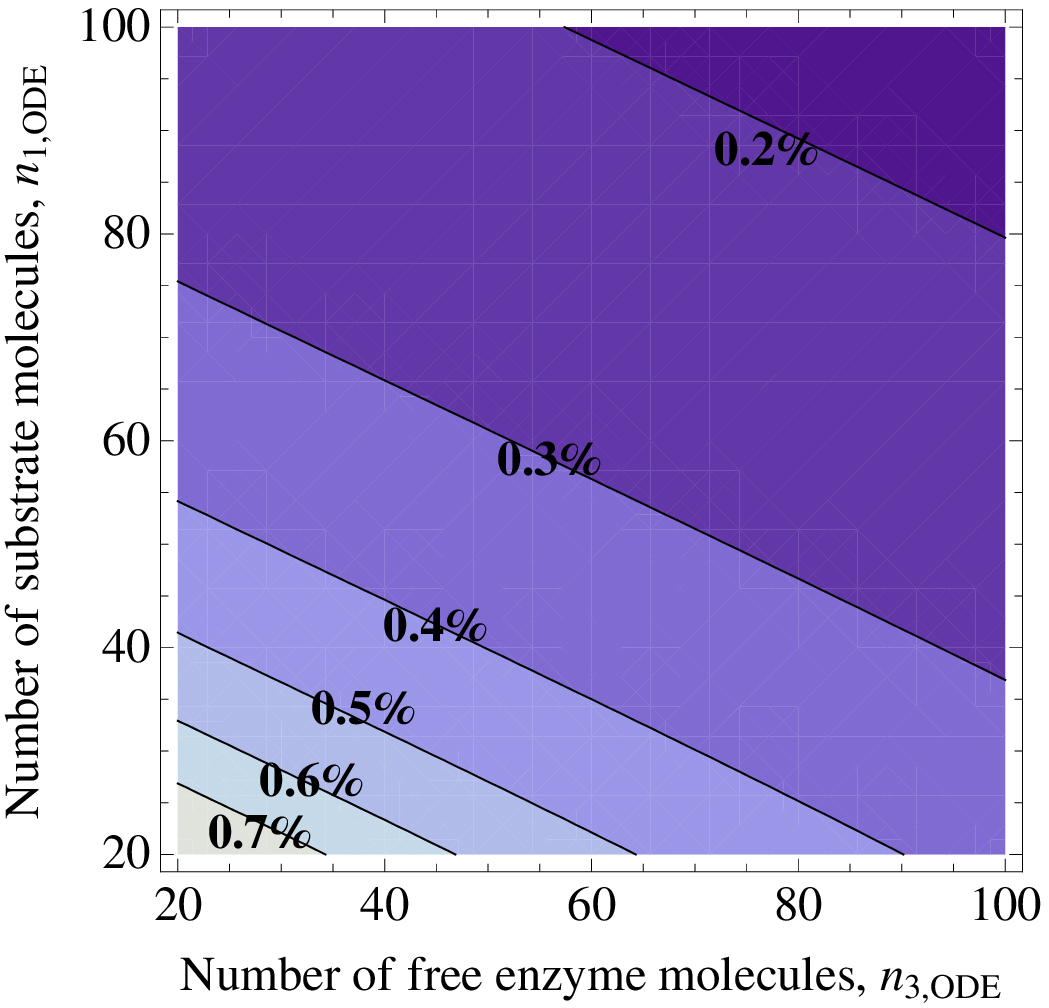}
\label{fig:subfig1}
}
\subfigure[]{
\includegraphics[width=3.0in]{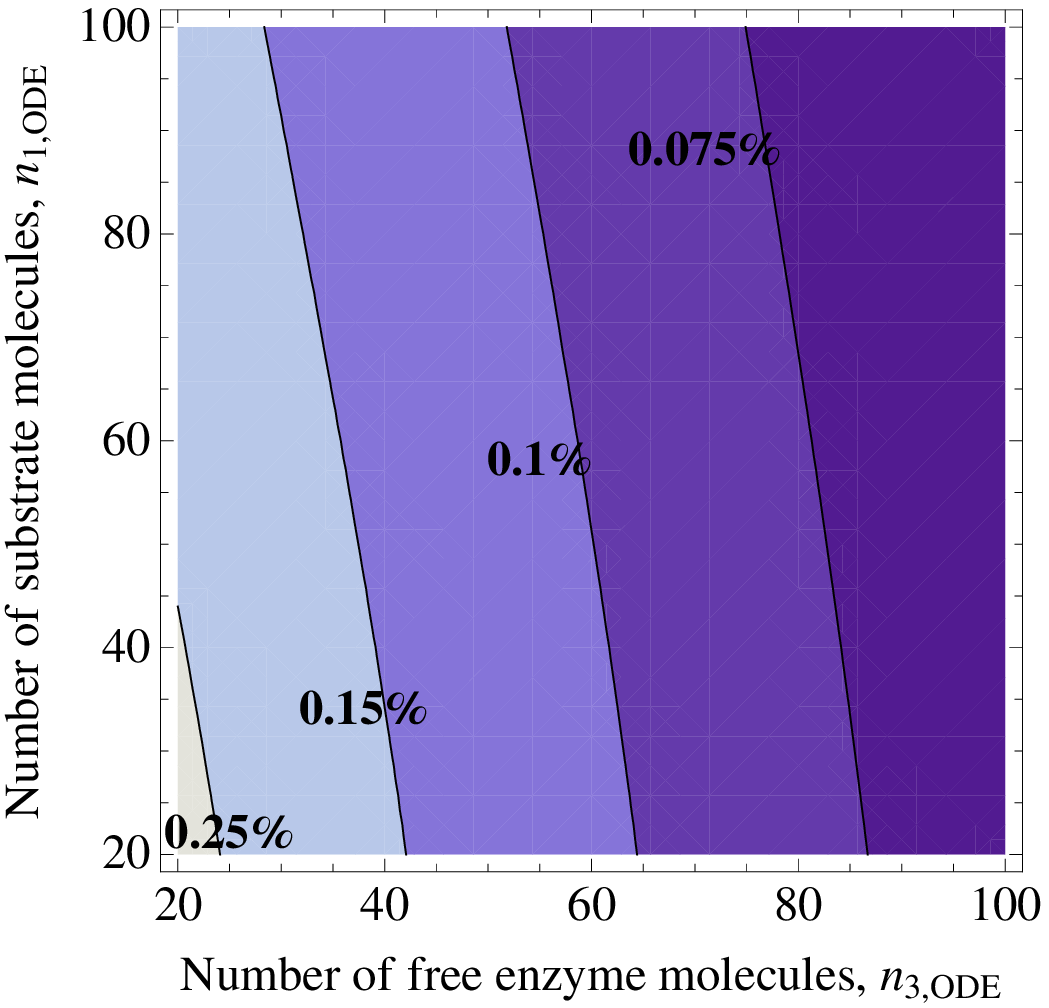}
\label{fig:subfig2}
}
\caption[]{Comparison of the predictions of the CFPE and CME for the Michaelis-Menten reaction mechanism. The differences between the two are quantified by calculation of the percentage relative error, i.e. 100 $\times$ (prediction of CME - prediction of CFPE) / prediction of CME. Panels (a) and (b) show  the maximum percentage relative error in the CFPE predictions of the variance of the substrate and complex concentration fluctuations, respectively. The figures are generated using Eqs. (94) and (95) together with Eq. (96); see text for details. The errors increase with decreasing molecule numbers; the magnitude of the error is very small in all cases implying that the CFPE is a highly accurate approximation of the CME.}
\end{figure}

\begin{figure}[ht]
\centering
\subfigure[]{
\includegraphics[width=3.0in]{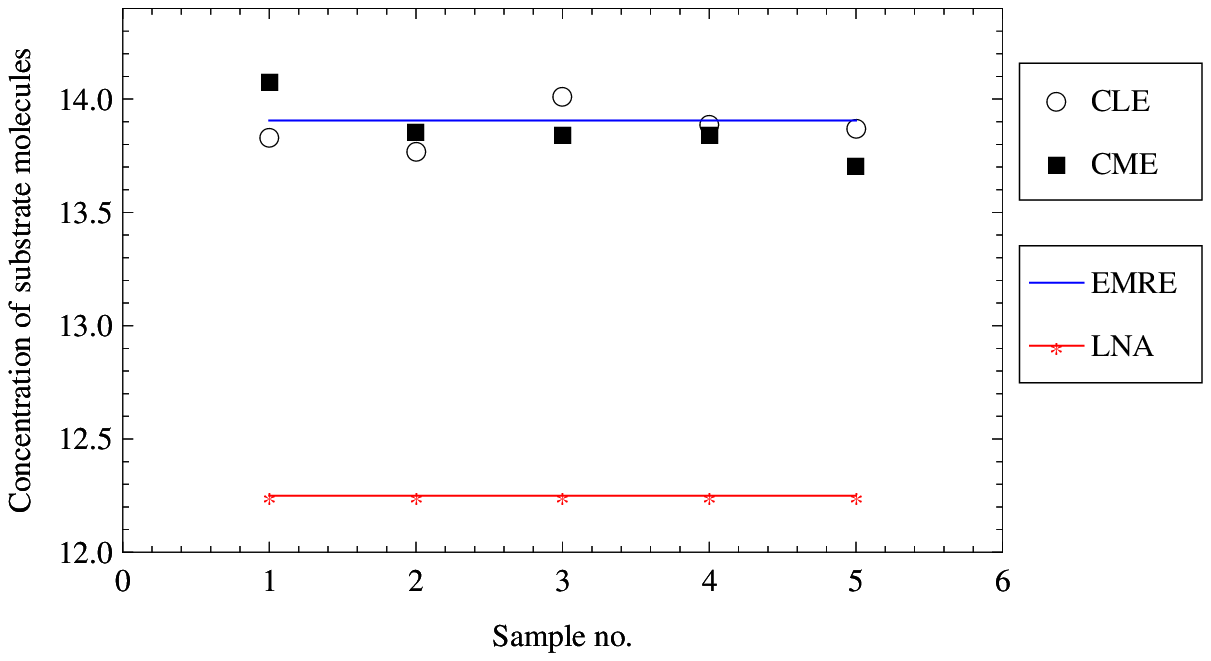}
\label{fig:subfig1}
}
\subfigure[]{
\includegraphics[width=3.0in]{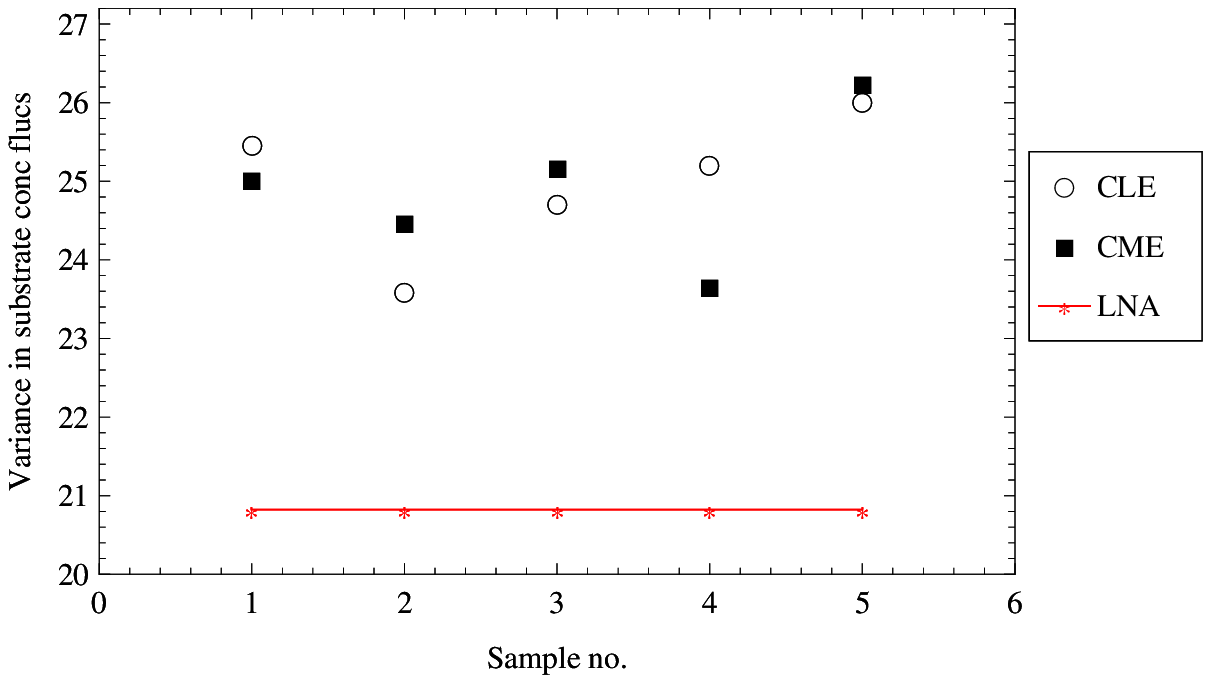}
\label{fig:subfig2}
}
\caption[]{Comparison of the predictions of the CLE for mean substrate concentration and variance of the fluctuations about the mean, with the predictions of the CME, the linear-noise approximation (LNA) and the mean concentration as predicted by Effective Mesoscopic Rate Equations (EMRE). Note that the CLE, within statistical error, is in agreement with the CME. The CLE predictions are more accurate than those obtained from the linear-noise approximation. The mean substrate concentration of the CLE agrees very well with the predictions of EMRE, Eq. (60).  See text for details.}
\end{figure}


\begin{thebibliography}{OHPHS}
\bibitem[35] {Samplingproc} The estimates were calculated according to the following procedure. Running the CLE Euler-Mayurama algorithm once leads to a set of numbers $\{  ( n_1(0), n_2 (0)),  (n_1(\Delta t), n_2 (\Delta t) ), (n_1(2\Delta t), n_2 (2\Delta t) ), ..... \} $ where $\Delta t$ is the discretization time step of the algorithm. An estimate of the steady-state average concentration of species $i$ as predicted by the CFPE is then given by $\langle n_i \rangle / \Omega = \Omega^{-1} \sum_{j=0}^{M} n_i(j \Delta t)$ where $M$ is the maximum number of time steps. Running the algorithm $N$ times leads to $N$ slightly different estimates of the mean of the CFPE; this variation leads to statistical error in the numerical estimation of the mean concentrations and can be reduced by increasing the total number of time steps. The same arguments hold for the estimation of the variance. For the simulations leading to Fig. 3 we chose $\Delta t = 5 \times 10^{-7}$ and $M = 10^{6}$. Smaller $\Delta t$ and larger $M$ did not change the results. Estimates of the concentrations and variances predicted by the CME can be obtained by sampling the trajectories of the stochastic simulation algorithm at intervals of time equal to the discretization time step of the CLE. The means and variance are then obtained as for the CLE.
\end{thebibliography}
\end{document}